\documentclass[aps, prx, twocolumn, showpacs, 10pt, floatfix, superscriptaddress, floatfix, longbibliography]{revtex4-2}
\usepackage[english]{babel}
\usepackage[utf8]{inputenc}
\usepackage[T1]{fontenc}

\usepackage{amsmath}
\usepackage{braket}
\usepackage{amssymb}
\usepackage[normalem]{ulem}
\usepackage{dsfont}
\usepackage{bm}
\usepackage[dvipsnames]{xcolor}
\usepackage{stmaryrd}
\usepackage{bbm}
\usepackage{mathtools}

\DeclarePairedDelimiter\floor{\lfloor}{\rfloor}

\usepackage{enumitem}
\usepackage{setspace}

\usepackage{color}
\usepackage[dvipsnames]{xcolor}

\usepackage[most]{tcolorbox}
\newtcbox{\othermathbox}[1][]{nobeforeafter, math upper, tcbox raise base, enhanced, rounded corners, colback=black!5, colframe=black, left=0.3em, top=0.3em, right=0.3em, bottom=0.4em}
\usepackage{empheq}
\usepackage[colorlinks,allcolors=blue]{hyperref}

\makeatletter
\def\l@subsection#1#2{}
\def\l@subsubsection#1#2{}
\makeatother

\usepackage{tikz}
\usetikzlibrary{decorations.text}

\newcommand{\beq}{\begin{equation}}
\newcommand{\eeq}{\end{equation}}
\newcommand{\phii}{\varphi}
\newcommand{\bA}{\textbf A}
\newcommand{\bB}{\textbf B}
\newcommand{\bF}{\textbf F}
\newcommand{\bj}{\textbf j}
\newcommand{\bJ}{\textbf J}
\newcommand{\bp}{\textbf p}

\newcommand{\bx}{\textbf x}

\newcommand{\cH}{H}
\newcommand{\cn}{n}
\newcommand{\cN}{{\cal N}}
\newcommand{\cO}{O}

\newcommand{\cV}{\mathcal{V}}
\newcommand{\der}{\partial}
\newcommand{\ds}{\displaystyle}

\newcommand{\eg}{{\it e.g.}\ }
\newcommand{\eps}{\varepsilon}

\DeclareMathOperator{\erfc}{erfc}

\newcommand{\ie}{{\it i.e.}\ }

\newcommand{\NN}{\mathbb{N}}

\newcommand{\RR}{\mathbb{R}}

\newcommand{\wick}[1]{\left. :\! \hspace{-0.5pt} #1 \hspace{-0.5pt} \!: \right.}
\newcommand{\pdag}{^{\vphantom{\dagger}}}
\newcommand{\ppr}{^{\vphantom{\prime}}}

\def\ii{\mathrm{i}}
\def\ee{\mathrm{e}}
\def\dd{\mathrm{d}}

\def\Ezo{E^0}
\def\Efo{E^1}

\renewcommand\i[1]{\textit{#1}}

\newcommand{\nocontentsline}[3]{}
\newcommand{\tocless}[2]{\bgroup\let\addcontentsline=\nocontentsline#1{#2}\egroup}

\begin{document}
%===============================================================

\title{Anisotropic Quantum Hall Droplets}

\author{Blagoje Oblak}
\email{blagoje.oblak@polytechnique.edu}
\affiliation{CPHT, CNRS, \'Ecole Polytechnique, IP Paris, F-91128 Palaiseau, France}
\author{Bastien Lapierre}
\email{bastien.lapierre@uzh.ch}
\affiliation{Department of Physics, University of Z\"urich, Winterthurerstrasse 190, 8057 Z\"urich, Switzerland}
\author{Per Moosavi}
\email{pmoosavi@phys.ethz.ch}
\affiliation{Institute for Theoretical Physics, ETH Zurich, Wolfgang-Pauli-Strasse 27, 8093 Z\"urich, Switzerland}
\author{Jean-Marie St\'ephan}
\email{stephan@math.univ-lyon1.fr}
\affiliation{\mbox{CNRS, Universit\'e Claude Bernard Lyon 1, UMR5208, Institut Camille Jordan, F-69622 Villeurbanne, France}}
\author{Benoit Estienne}
\email{estienne@lpthe.jussieu.fr}
\affiliation{Sorbonne Universit\'e, CNRS, Laboratoire de Physique Th\'eorique et Hautes Energies, LPTHE, F-75005 Paris, France}

\date{May 3, 2024}

\begin{abstract}
We study two-dimensional (2D) droplets of noninteracting electrons in a strong magnetic field, placed in a confining potential with arbitrary shape. Using semiclassical methods adapted to the lowest Landau level, we obtain near-Gaussian energy eigenstates that are localized on level curves of the potential and have a position-dependent height. This one-particle insight allows us to deduce explicit formulas for expectation values of local many-body observables, such as density and current, in the thermodynamic limit. In particular, correlations along the edge are long-ranged and inhomogeneous. As we show, this is consistent with the system's universal low-energy description as a free 1D chiral conformal field theory of edge modes, known from earlier works in simple geometries. A delicate interplay between radial and angular dependencies of eigenfunctions ultimately ensures that the theory is homogeneous in terms of the canonical angle variable of the potential, despite its apparent inhomogeneity in terms of more na\"ive angular coordinates. Finally, we propose a scheme to measure the anisotropy by subjecting the droplet to microwave radiation; we compute the corresponding absorption rate and show that it depends on the droplet's shape and the waves' polarization. These results, both local and global, are likely to be observable in solid-state systems or quantum simulators of 2D electron gases with a high degree of control on the confining potential.
\end{abstract}

%===============================================================
\maketitle
%===============================================================
\begin{spacing}{0}
\tableofcontents
\end{spacing}
%===============================================================

%===============================================================
\section{Introduction}
\label{Sec:Intro}
%===============================================================

Quantum Hall (QH) droplets are mesoscopic two-dimensional (2D) electron gases placed in a strong perpendicular magnetic field and confined by some electrostatic potential. They lie at the heart of the QH effect \cite{Klitzing, Tsui, LaughlinAnomalous} and provide a key benchmark for topological phases of matter as a whole. In practice, however, the majority of detailed analytical studies of QH droplets and their low-energy edge excitations \cite{Bahcall:1991an, Frohlich, Wen:1990qp, Wen:1990se, Wen, Boyarsky:2004ta} are limited to highly symmetric cases, typically involving isotropic traps or harmonic potentials that are translation invariant in one direction \cite{DelplaceMontambaux:2010, Montambaux:2011}. This is especially troubling as far as edge modes are concerned, since it is not obvious that they are universally described by a homogeneous 
\begin{figure}[ht]
\centering
\includegraphics[width=0.9\columnwidth]{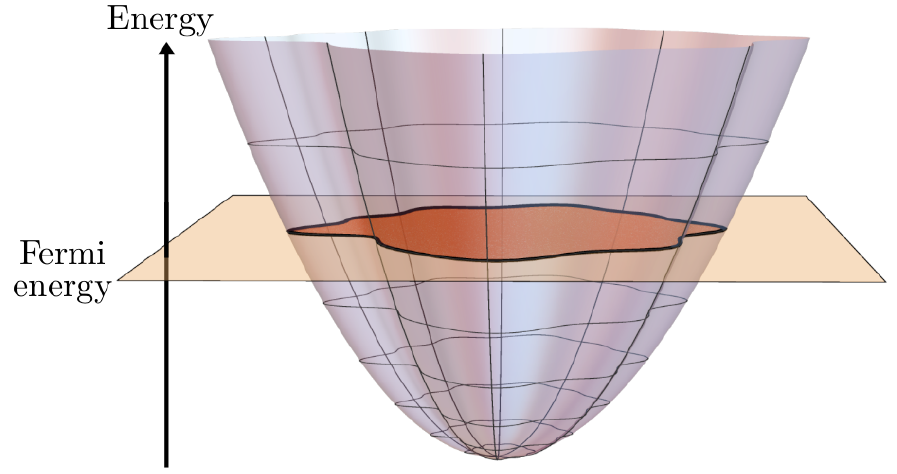}
\caption{2D electron droplet (shaded area) placed in a strong perpendicular magnetic field and confined by a typical anisotropic edge-deformed potential well \eqref{laviz}. At leading order in the thermodynamic limit, the droplet's boundary (thick black curve) coincides with the equipotential of the trap at the Fermi energy.}
\label{fiDrop}
\end{figure}
chiral Luttinger liquid when the gradient of the potential makes their propagation velocity position dependent \cite{Tomimatsu, Hotta:2022aiv, Kamiyama:2022fqhe}.

The goal of the present paper is to address this lack of analytical results by predicting the behavior of many-body observables near the edge of essentially \i{any} anisotropic droplet, as illustrated in Fig.~\ref{fiDrop}. We achieve this by providing general, explicit formulas for the density, current, and correlations in the regime of strong magnetic fields. We also study the corresponding low-energy edge modes, which are described by a free-fermion chiral conformal field theory (CFT) whose Fermi velocity is constant provided distances along the boundary are measured by the canonical angle coordinate determined by the potential. As we explain, this universal result stems from a delicate interplay between the radial and angular dependencies of anisotropic wave functions, and only becomes manifest upon suitably `averaging' over the radial direction.

Such phenomena are likely to be directly observable with local imaging techniques in condensed matter systems \cite{Weitz, Ahlswede, Weis, Ilani, Steele, Hashimoto2008, Hashimoto, BeckerEtAl:2011, Tomimatsu, Kamiyama:2022fqhe} or quantum simulators \cite{BretinEtAl:2004, Schweikhard, Cooper, Dellabetta, Goldman1, Goldman2, Fletcher, Mukherjee:2021jjl}. In addition, we predict two effects that probe the anisotropy without requiring local imaging. The first consists of shape-dependent quantum corrections to the velocity and dispersion of edge modes, measurable in both genuine QH samples \cite{AshooriEtAl:1992, KataokaEtAl:2016, HashisakaEtAl:2017, Maclure:2009} and their cold-atom simulators \cite{TaiEtAl:2017, LeonardEtAl:2023, BinantiEtAl:2023}; the second is the microwave absorption spectrum \cite{Cano} of anisotropic QH droplets, whose rich pattern of peaks with a polarization-dependent magnitude should similarly be detectable in view of the recent experiment \cite{Mahoney}.

Related questions have appeared in the literature. Indeed, random potentials with no symmetries are essential to model disorder, whose importance for the robustness of QH physics is hard to overstate \cite{Halperin, Trugman, Joynt}. A relevant series of works in that context is \cite{Champel1, Champel2}, which study the density and current of QH droplets with arbitrary potentials, at finite temperature, generally including Landau-level mixing, in the semiclassical limit of strong magnetic fields and weak traps \cite{Trugman, Joynt}. However, the coherent states used in these references only provide a limited resolution at the single-particle level, precluding the computation of low-energy dynamics and long-range correlations along the boundary. Our objective here is instead to find explicit one-particle wave functions, which will depend on the shape of the potential, and use these as a starting point for many-body objects.

Regarding electronic edge correlations, similar issues have been addressed in the context of classical 2D Coulomb gases, where holomorphic methods provide a handle on droplets of pretty much any shape \cite{ChoquardEtAl:1987, Jancovici:1995, ZabrodinWiegmann:2006, Klevtsov:2013iua, LebleSerfaty:2018, HedenmalmWennman:2021, AmeurCronvall:2023}. The most-studied case of harmonic traps even involves an exact correspondence between the quantum Landau problem and the appropriate Coulomb gas \cite{DiFrancesco:1994, ForresterJancovici:1996}. However, no such matching holds for generic confining potentials, so the two setups really need to be treated separately. In other words, the vast majority of anisotropic QH droplets admit no faithful Coulomb-gas description.

Finally, the results put forward here may be seen as microscopic, first-principles derivations of quantities that are normally studied within less-controlled approximation schemes in the geometry of the QH effect \cite{BradlynEtAl:2012ea, BradlynRead:2015, Ferrari:2014yba, Abanov:2014ula, Can1, Can2, Can3, Klevtsov:2015eda,Gromov:2015fda}. Our hope is thus to build a bridge between these theoretical works and concrete observations that may soon be accessible in tabletop experiments with a high degree of control on the confining potential \cite{Fletcher, Mukherjee:2021jjl}.

Here is the plan of the paper. To begin, Sec.~\ref{semeth} summarizes our methods and results, avoiding technical details. The next two sections are devoted to one-body physics in the lowest Landau level: Sec.~\ref{sesdiff} first discusses generalities on semiclassical holomorphic wave functions, and Sec.~\ref{sescale} presents a detailed calculation of the semiclassical energy spectrum in a broad class of `edge-deformed' potentials of particular interest. This leads to Sec.~\ref{semany}, where we investigate the many-body density, current, correlations and low-energy edge modes of anisotropic droplets. Last, Sec.~\ref{ARMA} is devoted to the microwave absorption spectrum, seen as a realistic global probe of anisotropy. We conclude in Sec.~\ref{seccc} by discussing several future directions and open questions. To streamline the text, some details are deferred to Appendices~\ref{appiso}--\ref{App:microwave_flower}.

%===============================================================
\section{Setup and main results}
\label{semeth}
%===============================================================

\begin{figure*}[t]
\centering
\includegraphics[width=\textwidth]{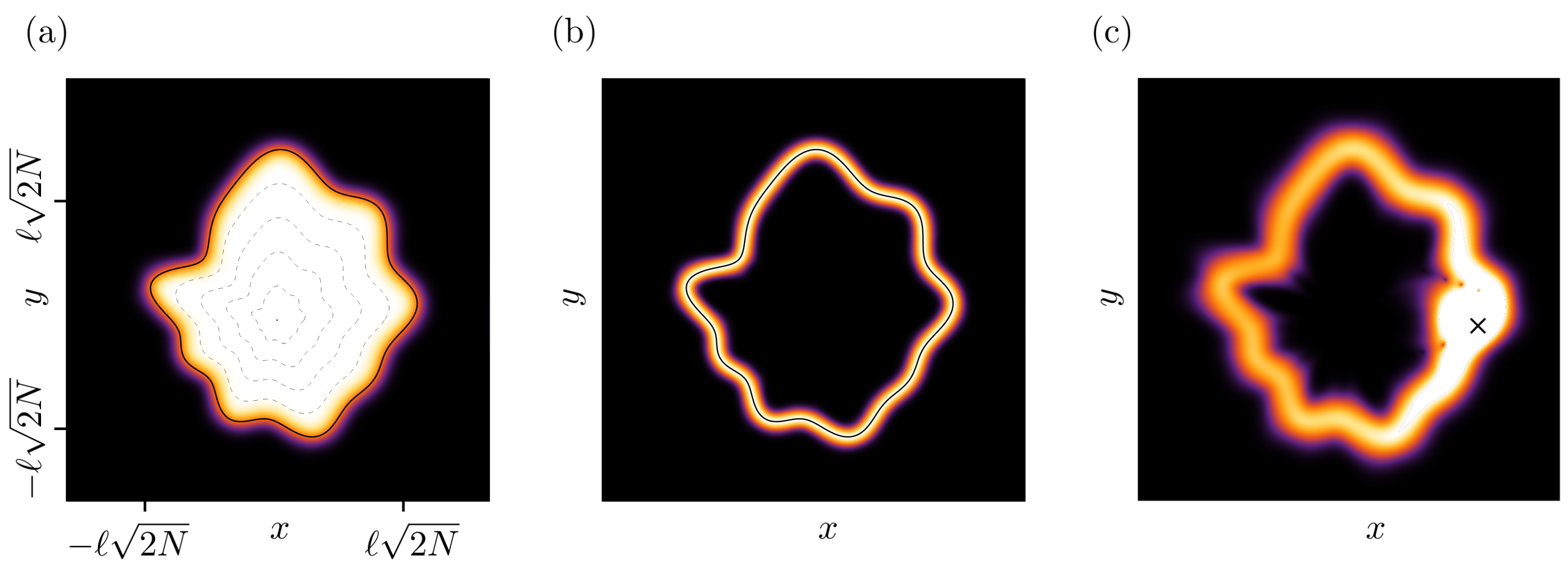}
\caption{Intensity plots of: (a) The many-body density \eqref{ressum_rho_edge} along with several equipotentials (dashed curves), for a droplet with $N=100$ electrons confined by the edge-deformed trap \eqref{laviz} used in Fig.~\ref{fiDrop}. The constancy of the bulk density and its decay at the boundary are manifest. (b) The norm of the current \eqref{ressum_J_edge} for the same droplet, together with the edge (black curve) on which it is localized. (c) The norm of the correlation function \eqref{cofo} for the same droplet, plotted as a function of $\bx_2=(x,y)$ for a fixed point (black cross) $\bx_1=(\ell\sqrt{N\lambda(0)},0)$ on the edge. Long-range correlations along the boundary are clearly visible and satisfy the asymptotics \eqref{ressum_C_edge} away from the coincident point $\bx_1=\bx_2$. In each case, the color coding goes from black to white, respectively corresponding to vanishing and maximal values of the plotted function.}
\label{fipott}
\end{figure*}

This section is an overview of our methods and results, beginning with the general setup (see Fig.~\ref{fiDrop}): a QH droplet in a strong magnetic field, with a trapping potential that varies slowly compared to the magnetic length \cite{Trugman, Joynt, GirvinLLL}. We explore this regime by developing a powerful WKB ansatz adapted to the lowest Landau level (LLL), inspired by semiclassical tools for holomorphic wave functions \cite{Zelditch, BleherEtAl:2000, MaMarinescu:book, ZelditchZhou} in general and quasimodes \cite{Charles:2003a, Charles:2003b} in particular. Concretely, we obtain the one-particle eigenfunctions and energy spectrum for a class of edge-deformed potentials representing the most general leading-order anisotropy of any star-shaped QH sample \cite{Note4}. We then apply these insights to the full many-body setting of an anisotropic QH droplet, providing explicit and practical formulas for both local and global many-body observables, respectively depicted in Figs.~\ref{fipott} and \ref{fiARMA}.

%---------------------------------------------------------------
\subsection{Semiclassical limit in the LLL}
%---------------------------------------------------------------

Consider spin-polarized noninteracting electrons of mass $M$ and charge $q$ in the plane. Each electron is governed by a Landau Hamiltonian with an anisotropic potential $V(\bx)$,
\beq
\label{h}
H_{\text{one-body}}
=
\frac{1}{2M}(\bp-q\bA)^2 + V(\bx),
\eeq
where $\bx$ denotes position, $\bp$ is canonical momentum, and $\bA$ is the vector potential of the magnetic field $\bB=\dd\bA$. The latter is taken to be uniform, \ie $\bB=B\,\dd x\wedge\dd y$ for some constant $B\neq0$ in terms of Cartesian coordinates $(x,y)$, and we systematically work in symmetric gauge $\bA=\tfrac{1}{2}B(x\,\dd y-y\,\dd x)$. (We view $\bA$ and $\bB$ as differential forms, which simplifies some notation but is otherwise inconsequential.) For simplicity, we also assume that $V(\bx)$ is `monotonic', meaning that it has a unique global minimum away from which it grows monotonically but is otherwise general \cite{Note:Reeb}. The level curves or \i{equipotentials} of $V(\bx)$ are therefore nested and take the form shown in Fig.~\ref{fipott}(a). Finally, we assume throughout that the potential is weak relative to the magnetic field \cite{Trugman, Joynt, Geller1, Geller2, Champel1, Champel2}, in that it is nearly constant on length scales comparable to the magnetic length $\ell$ given by
\beq
\label{ell}
\ell^2
\equiv
\frac{\hbar}{qB},
\eeq
where $qB>0$ without loss of generality \cite{Note1}. Note that the assumption of monotonicity is natural for QH samples: the potential near the edge of any realistic droplet is guaranteed to be monotonic, and this ultimately determines the low-energy physics regardless of bulk details.

In the regime of slowly varying potentials, the operator $V(\bx)$ is a small perturbation of the pure Landau Hamiltonian $\propto(\bp-q\bA)^2$ and the eigenstates of $H_{\text{one-body}}$ in Eq.~\eqref{h} are well approximated by wave functions in the LLL. For instance, if the potential $V(\bx) = V_{0}(r^2/2)$ is isotropic, any eigenfunction of $H_{\text{one-body}}$ has some definite angular momentum $\hbar m$ with integer $m$. Each eigenstate thus reduces at strong $B$ to a standard LLL wave function in symmetric gauge,
\beq
\label{phim}
\phi_m(\bx)
=
\frac{1}{\sqrt{2\pi\ell^2}}\,\frac{z^m}{\sqrt{m!}}\,\ee^{-|z|^2/2},
\eeq
where $m\geq0$ and we introduced the dimensionless complex coordinate
\beq
\label{zdef}
z
\equiv
\frac{x+\ii y}{\sqrt{2}\,\ell}.
\eeq
The wave function \eqref{phim} reaches its maximum on the circle $|z|=\sqrt{m}$, away from which it decays in a Gaussian manner with a width of the order of $\ell$. Our goal will be to obtain similar approximate eigenstates for \i{anisotropic} traps, using the squared magnetic length \eqref{ell} as a small parameter \cite{Note2}. Equivalently, we shall carry out a semiclassical (small $\hbar$), high-field (large $B$) expansion.

In practice, the projection to the LLL is implemented by the (one-body) operator $P\equiv \sum_{m=0}^{\infty}|\phi_m\rangle\langle\phi_m|$ whose kernel can be obtained from the wave functions \eqref{phim}:
\beq
\label{proj}
\langle z,\bar z|P|w,\bar w\rangle
=
\frac{1}{2\pi\ell^2}\,\ee^{-(|z|^2+|w|^2)/2}\,\ee^{z\bar w}.
\eeq
This kernel is manifestly Gaussian and reduces to a delta function in the formal semiclassical limit $\ell\to0$. At small but finite $\ell$, the projection \eqref{proj} makes space noncommutative in the sense that LLL-projected position operators satisfy the Heisenberg algebra
\beq
\label{pxp}
[PxP, PyP]
=
\ii\ell^2.
\eeq
One can thus think of the plane $\RR^2$ as a phase space whose canonical variables are $(x,y)$. This interpretation pervades much of the QH literature \cite{Bellissard, GirvinLLL, Girvin:1986zz, Pasquier:2007nda, Iso:1992aa, Susskind:2001fb, Polychronakos:2001mi, Hellerman:2001rj, Fradkin:2002qw, Dong:2020bkt, Du:2021hes} and will similarly affect our discussion. Indeed, projecting the Hamiltonian \eqref{h} to the LLL and looking for its spectrum leads to the eigenvalue equation
\beq
\label{claim}
PV\!P|\psi\rangle
=
E|\psi\rangle,
\eeq
where the unknowns are the energy $E$ and the quantum state $|\psi\rangle$ in the LLL \cite{Note3}. Note that the kinetic term in Eq.~\eqref{h} has disappeared in Eq.~\eqref{claim}: the potential itself plays the role of an effective Hamiltonian in the noncommutative phase space $(x,y)$.

Exact solutions of Eq.~\eqref{claim} are generally out of reach, so one has to resort to approximations. The semiclassical one that we shall use is well known in the QH context \cite{Trugman, Joynt, Champel1, Champel2, Charles:2003a, Charles:2003b}. More precisely, we will seek solutions of Eq.~\eqref{claim} labeled by a large quantum number $m\in\NN$, seen as a generalization of angular momentum. This large-$m$ limit is accompanied by a small-$\ell$ limit such that the area $2\pi\ell^2m$ remains fixed. In that regime, the $m^{\text{th}}$ eigenstate is approximately Gaussian and localized on an equipotential $\gamma_m$ of $V(\bx)$, enclosing a quantized area such that the Bohr-Sommerfeld condition holds:
\beq
\label{oint}
\oint_{\gamma_m} x\dd y
=
2\pi\ell^2\,m.
\eeq
Equivalently, the flux of the magnetic field through the area enclosed by $\gamma_m$ is $m$ times the flux quantum. The energy of the $m^{\text{th}}$ state is then
\beq
\label{ev}
E_m
=
\Ezo_m
+\ell^2\Efo_m
+\cO(\ell^4),
\eeq
where $\Ezo_m=V(\gamma_m)$ is the leading classical approximation and the quantum correction $\Efo_m$ involves the Laplacian of the potential and the curvature of the equipotential $\gamma_m$ \cite{Charles:2003a, Charles:2003b}. The more familiar Wentzel-Kramers-Brillouin (WKB) approximation of 1D quantum mechanics \cite{BenderOrszag:1999} includes (topological) Maslov corrections on the right-hand side of Eq.~\eqref{oint}; we will encounter similar corrections below, although their topological interpretation is prevented by a subtle distinction between real and K\"{a}hlerian polarizations in geometric quantization \cite{Charles:2003a, Charles:2003b}.

%---------------------------------------------------------------
\subsection{One-body results}
%---------------------------------------------------------------

The semiclassical limit just outlined applies to any (monotonic) weak potential. In practice, our main concern is the physics of QH droplets near the edge, where the details of the bulk potential are irrelevant. Most of our explicit results will therefore be given for edge-deformed potentials, obtained as follows. Consider any monotonically increasing function $V_{0}(s)$ for $s\geq0$, and let $\lambda(\phii)$ be any strictly positive $2\pi$-periodic function of the angle $\phii\in[0,2\pi)$. We normalize $\lambda(\phii)$ so that $\oint\dd\phii\,\lambda(\phii)=4\pi$, writing $\oint\dd\phii$ as a shorthand for $\int_0^{2\pi}\dd\phii$. Then, adopt polar coordinates in the plane such that $x+\ii y=r\ee^{\ii\phii}$ and define the potential
\beq
\label{laviz}
V(r,\phii)
\equiv
V_{0}\left(\frac{r^2}{\lambda(\phii)}\right).
\eeq
We refer to this as an \i{edge-deformed} trap because it arises from a deformation $r^2\mapsto r^2/\lambda(\phii)$ that changes the shape of the boundary of isotropic droplets in a finite and smooth way, even in the thermodynamic limit where the droplet's area goes to infinity \cite{EstienneOblakSDiff}. The corresponding equipotentials enclose star-shaped regions in the plane that only differ from one another by their overall scale \cite{Note4}. In this sense, the class of potentials \eqref{laviz} is generic as far as edge effects are concerned. It is partly inspired by earlier works on the $W_{1+\infty}$ algebra \cite{Cappelli1, Cappelli:1992kf, Cappelli:1993ei, Cappelli:1994wb, Cappelli:1995yk, Cappelli:1996pg, Cappelli:2021kxd}, where it was argued that infinitesimal deformations of the form \eqref{laviz} span a Virasoro algebra.

\begin{figure}[t]
\centering
\includegraphics[width=0.85\columnwidth]{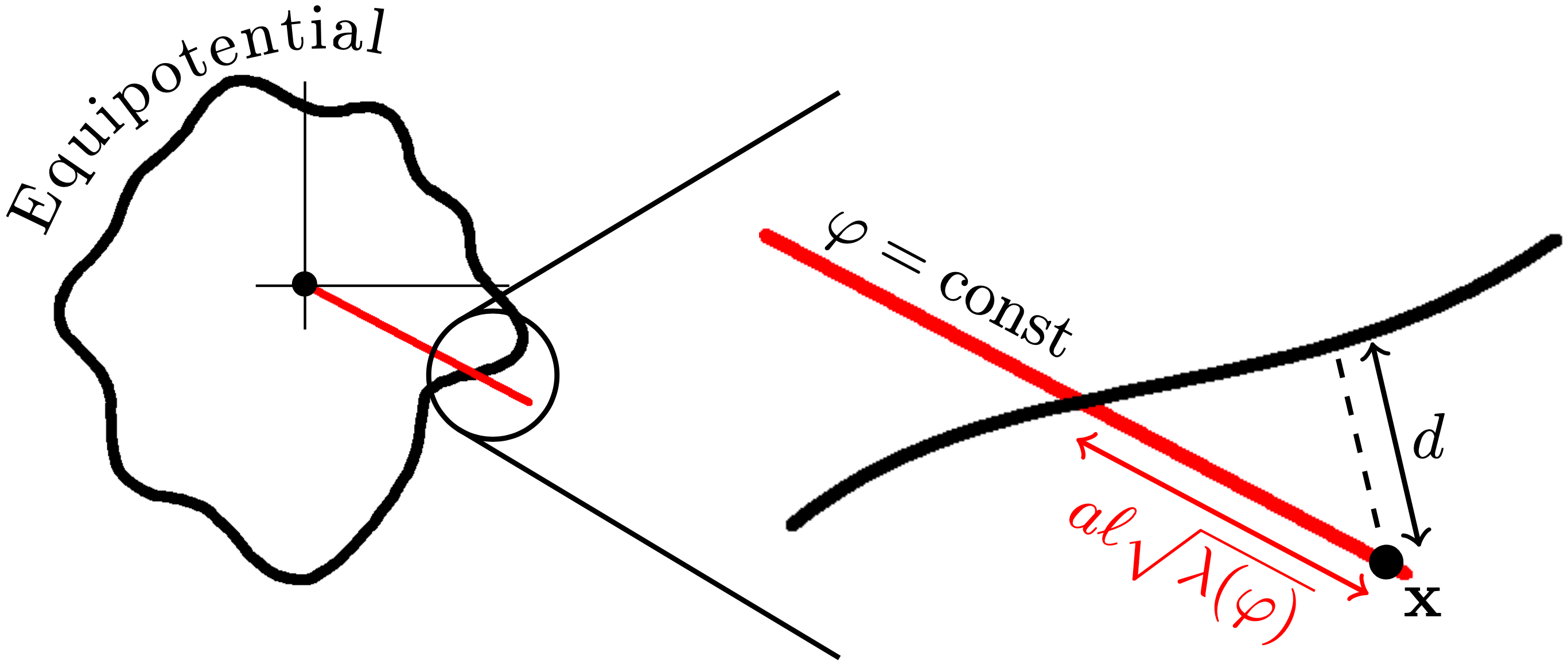}
\caption{The second exponential in Eq.~\eqref{e13} involves the signed distance $d$ between the point $\bx=(r\cos(\phii), r\sin(\phii))$ and the equipotential at $r = \ell\sqrt{m\lambda(\phii)}$. At large $m$, the equipotential is locally nearly straight, so the expression \eqref{edistance} of $d$ in terms of the angle $\phii$ and the radial deviation $a$ defined below Eq.~\eqref{e13} follows from elementary Cartesian geometry.}
\label{fidist}
\end{figure}

The traps \eqref{laviz} allow for explicit calculations of the semiclassical energy spectrum, generalizing the known isotropic formulas reviewed in Appendix~\ref{appiso}. Indeed, we show in Sec.~\ref{sescale} that the relevant eigenfunctions, solving the LLL eigenvalue problem \eqref{claim}, are Gaussians localized on equipotentials $r=\ell\sqrt{m\lambda(\phii)}$ at large quantum numbers $m$. They can be written in polar coordinates as
\beq
\label{e13}
\psi_m(r,\phii)
\sim
\frac{\ee^{\ii\Theta_m(r,\phii)}}{\sqrt{2\pi\ell^2 \sigma(\phii)}}\,
\frac{\ee^{-a^2/\sigma(\phii)^2}}{(2\pi m)^{1/4}},
\eeq
where $\Theta_m(r,\phii)$ is a position-dependent phase given below in Sec.~\ref{SubSec:GWF}, $a\equiv\bigl(r-\ell\sqrt{m\lambda(\phii)}\bigr)\big/\ell\sqrt{\lambda(\phii)}$  is a dimensionless coordinate encoding the deviation from the equipotential, and the quantity
\beq
\label{sigma_lambda}
\sigma(\phii)
\equiv
\sqrt{ \frac{2}{\lambda(\phii)} } \sqrt{1 + \left[ \frac{\lambda'(\phii)}{2\lambda(\phii)} \right]^2}
\eeq
determines the local height of the wave function. It is worth stressing the simple geometric interpretation of these objects. First, the ratio $a/\sigma(\phii)$ in the exponential in Eq.~\eqref{e13} measures the (signed) distance to the equipotential, namely
\beq
\label{edistance}
d = \sqrt{2}\,\ell \frac{a}{\sigma(\phii)}
\eeq
for large $m$; see Fig.~\ref{fidist}. The Gaussian factor in Eq.~\eqref{e13} thus exhibits the expected `quantum smearing' of wave functions in a strong but finite magnetic field \cite{Champel1, Champel2, ZelditchZhou}, which would be missed by the leading classical approximation ($\ell^2=0$). Second, the function \eqref{sigma_lambda} is proportional to the Euclidean norm of the velocity of guiding centers, namely
\beq
\|{\bf v}(\phii)\|
=
\sqrt{2m}\,
\frac{\ell V_{0}'(\ell^2m)}{qB}\,
\sigma(\phii)
\label{VELO}
\eeq
on the level curve $r=\ell\sqrt{m\lambda(\phii)}$ of an edge-deformed potential \eqref{laviz}. The probability density corresponding to the wave function \eqref{e13} is thus proportional to $1/\|{\bf v}(\phii)\|$, in agreement with classical intuition.

As for the energy of the state \eqref{e13}, its expansion \eqref{ev} up to $\cO(\ell^4)$ contributions turns out to be
\beq
\label{ressum_E_m}
E_m
\sim
V(\gamma_{m})
+
\frac{\ell^2}{2}\Omega_{m}
\biggl[1
+
\Bigl(1+\frac{\Gamma_{m}}{\Omega_{m}}\Bigr)
\oint\frac{\dd\phii}{4\pi}\lambda(\phii)\sigma(\phii)^2 \biggr],
\eeq
where $V(\gamma_{m}) = V_{0}\bigl(\ell^2m\bigr)$ is the leading (classical) term and the first quantum correction involves derivatives $\Omega_{m} \equiv V_{0}'(\ell^2m)>0$ and $\Gamma_{m} \equiv \ell^2m\, V_{0}''(\ell^2m)$. Note that our semiclassical regime ensures that $\Gamma_m/\Omega_m$ is finite at large $m$ \cite{Note5}. The $m^{\text{th}}$ energy is thus determined by the potential and its derivatives on an equipotential that satisfies the quantization condition \eqref{oint}, in accordance with general theorems for holomorphic WKB theory \cite{Charles:2003a, Charles:2003b}.

%---------------------------------------------------------------
\subsection{Many-body results}
%---------------------------------------------------------------

Now consider the ground state of a large number $N\gg1$ of free spin-polarized electrons, each governed by the single-particle Hamiltonian \eqref{h}. This ground state is a Slater determinant of wave functions whose large-$m$ behavior is the Gaussian \eqref{e13}. As we show in Secs.~\ref{semany} and~\ref{ARMA}, the ensuing many-body density, current, correlations, low-energy effective field theory, and microwave absorption spectrum can all be written in closed form in terms of $\lambda(\phii)$ and the number $N$ of fermions.

Concretely, the density $\rho(\bx)=\sum_{m=0}^{N-1}|\psi_m(\bx)|^2$ has the expected bulk value $\rho\sim\frac{1}{2\pi\ell^2}$, while its form near the edge at $r=\ell\sqrt{N\lambda(\phii)}$ is given by a complementary error function:
\beq
\label{ressum_rho_edge}
\rho(r,\phii)
\sim
\frac{1}{4\pi\ell^2}\,
\erfc \biggl( \frac{d(r, \phii)}{\ell} \biggr),
\eeq
where $d(r,\phii)$ is the (signed) distance to the droplet's edge \cite{ZelditchZhou}, given by Eq.~\eqref{edistance} for $m = N$. As a result, the ground state forms a star-shaped droplet with total area $2\pi N\ell^2$ and a nonzero width inherited from that of one-body wave functions. Turning to the current $\bJ=\sum_{m=0}^{N-1}\tfrac{1}{2\ii}(\psi_m^*\dd\psi_m-\psi_m\dd\psi_m^*-2\ii q|\psi_m|^2 \bA)$ written as a one-form in polar coordinates, one has
\beq
\label{ressum_J_edge}
\bJ(r, \phii)
\sim
-
\frac{\exp\Bigl(-\frac{d(r,\phii)^2}{\ell^2} \Bigr)}{(2\pi\ell^2)^{3/2} \sigma(\phii)}
\Bigl(
\ell\sqrt{N} \,\dd\phii + \frac{\lambda'(\phii)}{2\lambda(\phii)^{3/2}} \,\dd r\Bigr).
\eeq
This is localized on the edge and tangent to it, missing the bulk behavior $J_i\propto\varepsilon_{ij}\der_jV$ as expected in the LLL \cite{Geller1, Geller2}. Finally, the two-point correlation function
\beq
\label{cofo}
C(\bx_1,\bx_2)
=
\sum_{m=0}^{N-1}\psi_m^*(\bx_1)\psi_m(\bx_2)
\eeq
behaves near the edge as
\beq
\label{ressum_C_edge}
C(\bx_1,\bx_2)
\sim
\frac{\ee^{\ii\Theta_{N}(\bx_1,\bx_2)}}{4\pi\ell^2 \sqrt{\sigma(\phii_1)\sigma(\phii_2)}}
\frac{\ii\exp \left( -\frac{d_1^2}{2\ell^2}-\frac{d_2^2}{2\ell^2} \right)}
{\sqrt{2\pi N}\,\sin\bigl( \int_{\phii_2}^{\phii_1} \frac{\dd\phii}{4} \lambda(\phii) \bigr)}
\eeq
with $d_1 = d(|\bx_1|, \phii_1)$ and $d_2 = d(|\bx_2|, \phii_2)$ in polar coordinates, while $\Theta_{N}(\bx_1,\bx_2)$ is a complicated overall phase. Note again the Gaussian localization at the edge, as well as the long-range correlation $\propto\sin(\ldots)^{-1}$ typical of gapless fermions. Indeed, we will confirm that the underlying low-energy edge modes are described by a chiral CFT of free fermions; see the action functional \eqref{lactose} below. The corresponding angular Fermi velocity $\omega_{\text{F}}\sim\ell^2\Omega_N/\hbar$ [with $\Omega_N$ defined below Eq.~\eqref{ressum_E_m}] is constant along the boundary when measured in terms of the canonical angle variable of the potential \eqref{laviz}, namely
\beq
\label{theta}
\theta(\phii)\equiv\frac{1}{2}\int_0^{\phii} \dd\alpha\, \lambda(\alpha).
\eeq
By contrast, the `lab velocity' measured \eg in terms of Euclidean distances is generally nonconstant along the edge [recall Eq.~\eqref{VELO}].

\begin{figure}[t]
\centering
\includegraphics[width=0.8\columnwidth]{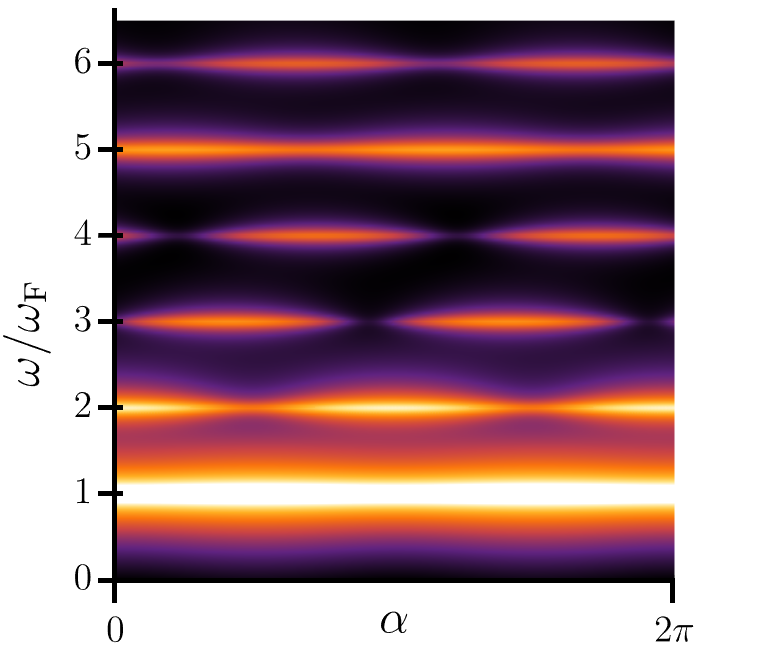}
\caption{Intensity plot of the absorption spectrum \eqref{79bis} for the same droplet as in Figs.~\ref{fiDrop} and \ref{fipott}, with delta functions replaced by Lorentzian distributions to account for the finite lifetime of quasiparticles in real systems. Following the dominant first peak at $\omega=\omega_{\text{F}}$ (the angular Fermi velocity), the absorption rate displays a series of weaker resonances at higher frequencies, with visibly angle-dependent magnitudes. The separation between peaks is clear despite the fact that the resolution chosen here is lower than what can be achieved in experiments \cite{Mahoney}.}
\label{fiARMA}
\end{figure}

The canonical angle coordinate \eqref{theta} crucially affects the microwave absorption spectrum. Indeed, we show in Sec.~\ref{ARMA} that an anisotropic droplet's electromagnetic absorption rate $\Gamma$ at frequency $\omega$ is given by
\begin{align}
\frac{\Gamma(\omega)}{2\pi N\ell^2}
&=
\frac{q^2E^2}{16\pi\hbar^2}
\sum_{p=1}^{\infty}p\,\delta(\omega-p\,\omega_{\text{F}}) \nonumber \\
&\quad\times\bigg|
\oint\frac{\dd\phii}{2\pi}\cos(\phii-\alpha)\,\ee^{\ii p\theta(\phii)}\,\lambda(\phii)^{3/2}
\bigg|^2, \label{79bis}
\end{align}
where the angle $\alpha$ determines the direction of the linearly polarized electric field with amplitude $E$. This predicts a series of peaks of absorption at resonance frequencies $p\,\omega_{\text{F}}$, as anticipated in \cite{Cano}; see Fig.~\ref{fiARMA} for a typical plot. For each absorption peak, both its height and its dependence on $\alpha$ involve the deformation function $\lambda(\phii)$ and the canonical angle \eqref{theta}. Eq.~\eqref{79bis} thus suggests that microwave absorption spectroscopy can be used to `hear the shape of a droplet'.

%===============================================================
\section{Anisotropic states from area-preserving deformations}
\label{sesdiff}
%===============================================================

This section presents the WKB ansatz [see Eq.~\eqref{final}] that forms the basis of all our later considerations. The structure is ultimately quite simple: given a monotonic potential $V(\bx)$, we pick one of its equipotentials, $\gamma_m$, with quantized area \eqref{oint}. We then build a wave function with winding $m$, perfectly localized on $\gamma_m$, and finally project it to the LLL using the operator \eqref{proj}. General theorems on K\"ahlerian semiclassical analysis \cite{BleherEtAl:2000, MaMarinescu:book, Zelditch, Charles:2003a, Charles:2003b} ensure that LLL-projected eigenstates satisfying Eq.~\eqref{claim} can indeed be built in this way. The detailed application of this method to edge-deformed traps \eqref{laviz} is given in Sec.~\ref{sescale}.

Note that what follows relies on the mathematics of area-preserving diffeomorphisms, which is not reviewed in detail. We refer instead to \cite{EstienneOblakSDiff} for an introduction whose language is similar to that adopted here. For more general discussions in the symplectic context, see \cite{IzKhMo:2016, DaSilva}.

%---------------------------------------------------------------
\subsection{Potentials in action-angle variables}
\label{ssepac}
%---------------------------------------------------------------

Let us be more precise about the geometry of the setup, remaining at the classical level for now.
We pick a smooth potential $V(\bx)$ and assume as in Sec.~\ref{semeth} that it is monotonic. Its unique global minimum is thus surrounded by nested level curves, and one can always find an area-preserving deformation of the plane that sends each equipotential on a circle \cite{IzKhMo:2016}. In other words, one can find an invertible smooth map $\bF:\RR^2\to\RR^2$ with unit Jacobian such that
\beq
\label{fivi}
V\big(\bF(\bx)\big)
=
V_{0}(r^2/2),
\eeq
where the trap on the right-hand side is isotropic, depending only on $r=|\bx|$. If $\bF$ is the identity (or a rotation around the origin), then $V$ was isotropic to begin with and its eigenstates satisfying Eq.~\eqref{claim} are the standard wave functions \eqref{phim} with definite angular momentum. In the more general case of arbitrary $V$, Eq.~\eqref{fivi} suggests using $\bF$ to map the eigenstates \eqref{phim} on those corresponding to our general $V(\bx)$.

The existence of $\bF$ in Eq.~\eqref{fivi} is guaranteed by the monotonicity and smoothness of $V$, and is equivalent to the existence of globally well-defined canonical action-angle variables. In fact, we can use this to write $\bF$ in a more explicit form that will be useful below. Let therefore $(\ell^2K,\theta)$ be action-angle coordinates for the potential $V(\bx)$, where $K\geq0$ is dimensionless and $\theta\in[0,2\pi)$ is a genuine angle. They are normalized so that $\ell^2\dd K\wedge\dd\theta=\dd x\wedge\dd y$, which is to say that their Poisson bracket reads $\{\ell^2K, \theta\}=\ell^2$ in terms of the phase space $(x,y)$ whose bracket stems from the commutator \eqref{pxp}. Then, the map $(x,y)\mapsto(\ell^2 K,\theta)$ is an area-preserving diffeomorphism in terms of which $V(\bx) = V_{0}(\ell^2K(\bx))$ is invariant under rotations of $\theta$. To be specific, write these coordinates as functions $K(x,y)$ and $\theta(x,y)$ and let the inverse be $x=F(K,\theta)$ and $y=G(K,\theta)$ for some functions $(F,G)$; this inverse is nothing but the deformation $\bF$ in Eq.~\eqref{fivi}. In other words, knowing the action-angle variables of a potential $V$ allows us to map it on its (unique) isotropic cousin $V_{0}$, which, in turn, can be used to relate the corresponding anisotropic eigenstates to those in Eq.~\eqref{phim}.

These considerations apply to any monotonic anisotropic trap, in which case one typically encounters intricate area-preserving maps with complicated action-angle variables. In practice, we will focus in Sec.~\ref{sescale} on the edge deformations mentioned below Eq.~\eqref{laviz}. This will enable us to tackle a broad range of setups where such complexities become manageable while still capturing key features of generic anisotropic droplets, especially as far as their boundary properties are concerned. For now, we remain general and turn to quantum aspects.

%---------------------------------------------------------------
\subsection{Anisotropic eigenstates}
\label{sseniseg}
%---------------------------------------------------------------

Using the action-angle variables $(\ell^2K,\theta)$ for $V(\bx)$, the statements around Eqs.~\eqref{oint}--\eqref{ev} can be turned into formulas and eventually yield anisotropic eigenfunctions that satisfy Eq.~\eqref{claim}. Indeed, the Bohr-Sommerfeld quantization condition \eqref{oint} implies that the equipotential $\gamma_m$ is the set of points in $\RR^2$ for which $K = m$. Now consider the following quantum state, perfectly localized on $\gamma_m$:
\beq
\label{psim}
|\Psi_m\rangle
\equiv
2\pi\ell^2
\oint\dd\theta\,
\cn(\theta)
\,\ee^{\ii m\theta} \big|F(m,\theta),G(m,\theta)\big\rangle.
\eeq
Here the factor $2\pi\ell^2$ is included for later convenience, the state $|F,G\rangle$ gives a delta function $\langle\bx|F(m,\theta),G(m,\theta)\rangle=\delta^2\big(\bx-\bF(m,\theta)\big)$, and $\cn(\theta)$ is some complex periodic function. The latter 
does not wind upon completing one turn in the plane along the equipotential, so all the winding of the integrand in Eq.~\eqref{psim} is encoded in the phase $\ee^{\ii m\theta}$.

We emphasize that Eq.~\eqref{psim} is analogous to the standard WKB ansatz $\psi(x)\sim\ee^{\ii S_0(x)/\hbar}\ee^{\ii S_1(x)}$ in 1D. Indeed, the phase $\ee^{\ii m\theta}$ is the leading classical contribution $\ee^{\ii S_0/\hbar}$ for $m \gg 1$, corresponding to the geometrical-optics approximation of the wave function, while $\cn(\theta)$ is the physical-optics quantum correction $\ee^{\ii S_1}$ that needs to satisfy a transport equation in order for the Schr\"odinger equation to hold \cite{BenderOrszag:1999}. The only difference lies in the interpretation of areas in the plane as values of an `action', which ultimately stems from the noncommutative geometry \eqref{pxp} of LLL physics. Note that $\cn(\theta)$ is the only unknown in Eq.~\eqref{psim}; in fact, most of the WKB method below will concern the derivation of a transport equation for $\cn(\theta)$ from the requirement that Eq.~\eqref{claim} be satisfied.

Starting from Eq.~\eqref{psim}, it is straightforward to build a state in the LLL thanks to the projector \eqref{proj}: denoting
\beq
\label{e32}
\psi_m(z,\bar z)
\equiv
\langle z,\bar z|P|\Psi_m\rangle,
\eeq
one finds the wave function
\begin{multline}
\label{final}
\psi_m(z,\bar z)
=
\ee^{-|z|^2/2}
\oint\dd\theta\,\cn(\theta)\,\ee^{\ii m\theta} \qquad\quad \\
\times \ee^{-[F(m,\theta)^2 + G(m,\theta)^2] / 4\ell^2}
\ee^{z[F(m,\theta) - \ii G(m,\theta)] / \sqrt{2}\ell}.
\end{multline}
This is manifestly of the form $\ee^{-|z|^2/2}$ times a holomorphic function that depends on the action variable $\ell^2m$ and the uniformizing map $\bF$ in Eq.~\eqref{fivi}. It will be our starting point for the semiclassical solution of the eigenvalue equation \eqref{claim}.

As a consistency check, note that Eq.~\eqref{final} simplifies for isotropic potentials. In that case, the action-angle variables are essentially polar coordinates $\ell^2K=r^2/2$ and $\theta=\phii$, and the map in Eq.~\eqref{fivi} is $\bF(\bx)=\bx$, merely implementing a change from polar to Cartesian coordinates: $F(m,\theta)=\ell\sqrt{2m}\cos(\theta)$ and $G(m,\theta)=\ell\sqrt{2m}\sin(\theta)$. One can then verify that $\psi_m(z,\bar z)$ in Eq.~\eqref{final} with $\cn(\theta)=\text{const}$ coincides (up to normalization) with the standard LLL wave function \eqref{phim}. Similarly to that simple example, any anisotropic wave function \eqref{final} reaches its maximum on the equipotential $\gamma_m$ and is approximately Gaussian close to it, as ensured by the kernel \eqref{proj}. This will be confirmed explicitly in Sec.~\ref{sescale} for edge deformations.

%---------------------------------------------------------------
\subsection{Expanding the eigenvalue equation}
\label{ssexp}
%---------------------------------------------------------------

None of what we wrote so far involves a manifest semiclassical expansion; the latter is hidden in the eigenvalue equation \eqref{claim} and the function $\cn(\theta)$ in Eq.~\eqref{final}, since $\cn(\theta)$ should be expanded as a power series $\cn(\theta) = \cn_0(\theta) + \ell^2 \cn_1(\theta) + \cO(\ell^4)$. (As before, there are no odd powers of $\ell$ since $\ell^2 \propto \hbar$ is really the semiclassical parameter.) It is therefore worth anticipating the first few terms of the semiclassical approximation of Eq.~\eqref{claim}. We stress that the expansion below will eventually be limited to the leading-order transport equation, so only $\cn_0(\theta)$ will, in fact, be calculated. In principle, one can push the expansion to higher orders for more detailed results.

The semiclassical expansion of the right-hand side of Eq.~\eqref{claim} is clear: it is given by the large-$m$, small-$\ell^2$ expansion of the projected wave function \eqref{final}, including an expansion of $\cn(\theta)$.
As for the energy, its expansion was written in Eq.~\eqref{ev}. The left-hand side of Eq.~\eqref{claim} is more subtle, as its semiclassical expansion involves that of the operator $PV\!P$. The latter is a `Berezin-Toeplitz operator' \cite{Charles:2003a, Charles:2003b} that will play an important role for edge-deformed potentials, so we now explain its expansion in some detail. First, given Cartesian coordinates $(x,y)$, express the potential in complex coordinates \eqref{zdef} as $V(x,y) \equiv \cV(z,\bar z)$ for some function $\cV(z,\bar{w})$ which is holomorphic in $z$ and antiholomorphic in $w$. Then, recall that $P$ is the LLL projector with kernel \eqref{proj} to find
\begin{multline}
\label{interm}
\langle z,\bar{z}|
PV\!P
|w,\bar{w}\rangle
=
\frac{1}{2\pi\ell^2}\,
\ee^{-(|z|^2+|w|^2)/2} \\
\times
\int_{\RR^2}\dd u \dd v\,V(u,v)\,
\ee^{-|X|^2+z\bar{X}+\bar{w}X}
\end{multline}
with $X\equiv(u+\ii v)/\sqrt{2}\ell$ defined similarly to the coordinate \eqref{zdef}. Our task is to expand the integral on the right-hand side in the semiclassical limit. The key is to assume that the potential varies slowly on the scale of the magnetic length \cite{Trugman, Joynt, Champel1, Champel2}, \ie we choose once and for all a smooth potential $V(\bx)$, independent of $\ell$, and let $\ell$ be small. In that regime, the integrals in Eq.~\eqref{interm} are approximately Gaussian and give (see Appendix~\ref{App:PVP_expansion})
\begin{multline}
\label{topop}
\langle z,\bar z|
PV\!P
|w,\bar w\rangle
\stackrel{\ell \ll 1}{\sim}
\frac{1}{2\pi\ell^2}\,
\ee^{-|z-w|^2/2}\,
\ee^{(z\bar w-\bar zw)/2} \\
\times
\left[\cV(z,\bar w)+\frac{\ell^2}{2}(\nabla^2V)(z,\bar w)\right],
\end{multline}
where $(\nabla^2V)(z,\bar w)$ is the bicomplex function that corresponds to the Laplacian of the potential, \ie $(\nabla^2V)(z,\bar w) = \tfrac{4}{2\ell^2} \partial_z \partial_{\bar{w}}\cV$. This is the standard semiclassical expansion of a Berezin-Toeplitz operator \cite{BleherEtAl:2000,Charles:2003a, Charles:2003b}. Note the general structure: the entire $PV\!P$ operator boils down to $P$ itself, with kernel \eqref{proj}, multiplied by a function that coincides with $V$ at leading order but also includes quantum corrections. In the `zoomed-out' limit where the kernel of $P$ is a delta function, the first term in Eq.~\eqref{topop} becomes $\cV(z,\bar z)\delta^2(z-w,\bar z-\bar w)$ as expected. Moreover, for harmonic potentials, the truncated expression \eqref{topop} is actually \i{exact} since the next term $\nabla^4 \cV$ and all subsequent ones vanish. This agrees with the common lore that `WKB is exact for quadratic Hamiltonians'.

%===============================================================
\section{Edge-deformed anisotropic traps}
\label{sescale}
%===============================================================

Here, we apply the WKB ansatz of Sec.~\ref{sesdiff} to potentials \eqref{laviz} with scale-invariant level curves, obtained by acting with edge deformations on an isotropic trap. As we explain below, these provide a broad class of setups that describe QH droplets whose shape is any star domain in the plane \cite{Note4}. The plan is as follows. First, we introduce edge deformations and give a few examples for later reference. Second, we apply Eq.~\eqref{claim} to edge-deformed traps and expand it in the classical limit [large $m$, small $\ell^2$ with $\ell^2m = \cO(1)$ kept fixed]. We keep track of all terms up to order $\cO(\ell^2)$, leading to a transport equation for the function $\cn(\theta)$ in Eqs.~\eqref{psim} and \eqref{final}. This eventually yields an explicit energy spectrum [see Eq.~\eqref{emi}] along with approximately Gaussian eigenfunctions [see Eq.~\eqref{gaussbis}]. Last, we conclude with a consistency check by showing that our wave functions reproduce the asymptotic (large-$m$) form of the known LLL-projected spectrum for anisotropic harmonic traps \cite{Jancovici:1981, ForresterJancovici:1996, Rider:2003, Bender:2010, LeeRiser:2016}.

%---------------------------------------------------------------
\subsection{Edge deformations}
%---------------------------------------------------------------

We saw in Sec.~\ref{sesdiff} that area-preserving deformations play a key role for the semiclassical solution of the eigenvalue equation \eqref{claim}. The group of all such deformations is obviously huge, so it is essential to identify the subset of transformations that are likely to be important for low-energy physics. In fact, part of this work has already been carried out, at least implicitly, in \cite{Cappelli1, Cappelli:1992kf, Cappelli:1993ei, Cappelli:1994wb, Cappelli:1995yk, Cappelli:1996pg}, which we now use as a basis for the definition of edge deformations. (A similar motivation was put forward in \cite{EstienneOblakSDiff}.)

Label points on the plane by their polar coordinates $(r, \phii)$, defined as usual by $x + \ii y = r\,\ee^{\ii\phii}$. Then, the boundary of any isotropic QH droplet is located at some fixed radius $r_{\text{edge}} = \cO(\ell\sqrt{N})$. What is the most general area-preserving diffeomorphism that preserves this order of magnitude? The answer is readily found by realizing that the constraint of keeping $r_{\text{edge}} = \cO(\ell\sqrt{N})$ is equivalent, at leading order in $1/N$, to the condition that the deformation commutes with overall dilations $r \mapsto \mathrm{const} \times r$. The most general diffeomorphism satisfying this criterion is an \i{edge deformation}
\beq
\label{edd}
\bigg(\frac{r^2}{2},\phii\bigg)
\mapsto
\bigg(\frac{r^2}{2f'(\phii)},f(\phii)\bigg),
\eeq
where $f(\phii)$ is an (orientation-preserving) deformation of the circle, \ie any smooth map satisfying $f(\phii+2\pi)=f(\phii)+2\pi$ and $f'(\phii)>0$ \cite{Note7}. The angle-dependent rescaling of $r$ on the right-hand side ensures that the map preserves area. Note that the set of maps \eqref{edd} is isomorphic to the group of diffeomorphisms of the circle, whose central extension famously leads to the Virasoro algebra encountered in CFT. Indeed, this motivates the statement in \cite{Cappelli:1993ei, Cappelli:1994wb} that generators of maps \eqref{edd} in the QH effect produce conformal transformations of edge modes.

We stress that the subset of transformations \eqref{edd} originates from an asymptotic analysis of the relevant orders of magnitude. One can undoubtedly consider other families of deformations, motivated by different considerations, but those are irrelevant for our purposes. For instance, the transformations $r^2\mapsto r^2+\alpha(\phii)$ are crucial for the effective low-energy description of QH droplets \cite{Wen, Bahcall:1991an, Cappelli:1994wb}, but they are subleading compared to those in Eq.~\eqref{edd} since they deform the radius $r_{\text{edge}}=\cO(\ell\sqrt{N})$ by terms of order $\cO(1/N)$ instead of $\cO(1)$. Conversely, one might consider `higher-spin transformations' \cite{Cappelli1, Cappelli:1993ei, Cappelli:1994wb} that change the radius in a dramatic way such as $r^2\mapsto\beta(\phii)r^4[1 + \cO(1/r)]$, but these stretch QH droplets to an infinite extent in the thermodynamic limit, which is why we discard them.

Let us provide a few examples of edge deformations for future reference. First, the maps \eqref{edd} include rigid rotations around the origin given by $f(\phii) = \phii + \mathrm{const}$. A richer class is obtained by fixing some positive integer $k$ and considering all maps of the form
\beq
\label{boost}
\ee^{\ii kf(\phii)}
=
\frac{\alpha\,\ee^{\ii k\phii}+\beta}{\bar\beta\,\ee^{\ii k\phii}+\bar\alpha},
\eeq
where $\alpha,\beta$ are complex numbers satisfying $|\alpha|^2-|\beta|^2 = 1$. For fixed $k$, such maps span a group locally isomorphic to $\mathrm{SL}(2,\RR)$, always containing a subgroup of rigid rotations. We will return to these deformations below, since they can be seen as Fourier modes for circle diffeomorphisms \cite{Note:Fourier}. In particular, setting $\alpha = \cosh(\lambda)$ and $\beta = \sinh(\lambda)$ for some real parameter $\lambda$ turns the map \eqref{boost} into an analogue of a Lorentz boost with rapidity $\lambda$. In terms of the bulk action \eqref{edd}, any deformation \eqref{boost} turns a circle into a `flower with $k$ petals'; see Fig.~\ref{fidens} for $k=3$. For $k = 2$, this maps the circle on an ellipse (see \cite{EstienneOblakSDiff} for details), to which we will return in Sec.~\ref{SubSec:Elliptic_comparison}.

%---------------------------------------------------------------
\subsection{Edge-deformed potentials}
%---------------------------------------------------------------

Given an isotropic potential $V_{0}(r^2/2)$, how is it affected by an edge deformation \eqref{edd}? The answer is provided by the anisotropic trap \eqref{laviz} with $\lambda(\phii) = 2f'(\phii)$:
\beq
\label{viviz}
V(r,\phii)
\equiv
V_{0}\biggl(\frac{r^2}{2f'(\phii)}\biggr).
\eeq
In what follows, we exclusively consider this class of potentials and refer to them as edge-deformed traps, for the reasons stated above. The shape of their equipotentials is entirely fixed by the function $f(\phii)$. For instance, flower deformations \eqref{boost} give rise to lower values of the potential inside the flower's petals and higher values between petals. Having chosen some circle deformation $f(\phii)$, our goal is to solve the corresponding eigenvalue equation \eqref{claim} in the classical limit of high quantum numbers and small magnetic length.

We begin by listing the key classical data of the problem. The action-angle variables suited to Eq.~\eqref{viviz} are $(\ell^2K, \theta) = \bigl(r^2\big/2f'(\phii), f(\phii) \bigr)$ with an inverse given by $(r^2/2,\phii) = \big(\ell^2K/(f^{-1})'(\theta),f^{-1}(\theta)\big)$, where $f^{-1}$ denotes the inverse of $f$. In particular, the angle coordinate coincides with our earlier Eq.~\eqref{theta} upon using $\lambda(\phii)=2f'(\phii)$, possibly up to an overall rotation of $\theta$. Points satisfying
\beq
\label{epot}
\frac{r^2}{2f'(\phii)}=\ell^2 K
\eeq
with constant $K\geq0$ form an equipotential, \ie a level curve of the potential in Eq.~\eqref{viviz}. In Cartesian coordinates, this is the set of points $x = \sqrt{2\ell^2K f'(\phii)} \cos(\phii)$, $y = \sqrt{2\ell^2K f'(\phii)} \sin(\phii)$ for $\phii \in [0, 2\pi)$. Equivalently, in terms of the angle variable $\theta=f(\phii)\in [0, 2\pi)$, the equipotential is
\beq
\label{xypott}
\begin{aligned}
x
& = \sqrt{\frac{2\ell^2K}{(f^{-1})'(\theta)}} \cos(f^{-1}(\theta))\equiv F(K,\theta),\\
y
& = \sqrt{\frac{2\ell^2K}{(f^{-1})'(\theta)}} \sin(f^{-1}(\theta))\equiv G(K,\theta),
\end{aligned}
\eeq
where the notation $(F,G)$ was introduced in Sec.~\ref{ssepac}. We will eventually focus on the regime where $K$ is a very large integer $m$ such that the dimensionful area $2\pi\ell^2 m$ is an $\cO(1)$ quantity as $\ell\to0$.

Moving just slightly away from the classical regime, we saw in Sec.~\ref{sesdiff} that the expansion of the operator $PV\!P$ involves a bicomplex potential function $\cV(z,\bar w)$. In the case of edge-deformed potentials \eqref{viviz}, with the convention \eqref{zdef} for complex coordinates, one finds
\beq
\label{vzw}
\cV(z,\bar w)
=
V_{0}\left( \ell^2\frac{z\bar w}{f'\big(\tfrac{1}{2\ii}\log[z/\bar w]\big)} \right).
\eeq
Note that this only makes sense for $z$ and $w$ close to each other; otherwise, taking $z \to \ee^{2\pi\ii}z$ affects the argument of $f'$ on the right-hand side. By contrast, when $z$ and $w$ remain close, taking $z \to \ee^{2\pi\ii}z$ also requires $w \to \ee^{2\pi\ii}w$, and this time the angle $\tfrac{1}{2\ii}\log[z/\bar w]$ is indeed invariant.

Finally, the expansion \eqref{topop} also involves the complexified Laplacian of the potential, but only its real value will be relevant at the order studied here. Let us therefore express the Laplacian of Eq.~\eqref{viviz} in polar coordinates:
\begin{align}
\nabla^2V
& =
\frac{1}{f'}
\left(2-\frac{1}{2}\frac{f'''}{f'}+\frac{f''^2}{f'^2}\right) V_{0}'\left(r^2/2f'\right) \nonumber \\
& \quad +
\frac{r^2}{f'^2}\left(1+\frac{f''^2}{4f'^2}\right) V_{0}''\left(r^2/2f'\right).
  \label{lav}
\end{align}
Here, the prime means differentiation with respect to the argument, namely $\phii$ for $f(\phii)$ and $r^2/2$ for $V_{0}(r^2/2)$. We shall rely on Eqs.~\eqref{vzw} and \eqref{lav} below, since they directly affect the eigenvalue equation \eqref{claim}.

%---------------------------------------------------------------
\subsection{Eigenvalue equation and energy}
\label{ssegeqen}
%---------------------------------------------------------------

Having studied the potential \eqref{viviz}, let us turn to the quantum state meant to solve the eigenvalue equation \eqref{claim}. As in Sec.~\ref{sseniseg}, we begin by building a state \eqref{psim} that is perfectly localized on the equipotential \eqref{epot} with $K=m$, project to the LLL using the operator \eqref{proj}, and obtain the wave function \eqref{final} that now reads
\begin{multline}
\label{true}
\psi_m(z,\bar z)
=
\ee^{-|z|^2/2}
\oint \dd\phii\, f'(\phii)\, \cn(f(\phii)) \\
\times \exp\left[ \ii mf(\phii)-\frac{1}{2}mf'(\phii)+z\,\sqrt{mf'(\phii)}\,\ee^{-\ii\phii} \right],
\end{multline}
where we changed variables using $\theta = f(\phii)$. It remains to show that this solves the eigenvalue equation \eqref{claim} for edge-deformed traps \eqref{viviz} in the semiclassical regime, provided the function $\cn(\theta)$ satisfies a suitable transport equation. The latter is derived by expanding the energy \eqref{ev} and the potential \eqref{topop} to get
\begin{align}
0
& =
\oint \dd\phii\, f'(\phii)\, \cn(f(\phii)) \nonumber \\
& \quad \times
\biggl(\cV(z,\bar w)+\frac{\ell^2}{2}\nabla^2V(z,\bar w)-\Ezo_m-\ell^2\Efo_m\biggr) \nonumber \\
& \quad \times
\exp\biggl[\ii mf(\phii)-\frac{1}{2}mf'(\phii)+z\bar w\biggr]\bigg|_{\bar w=\sqrt{mf'(\phii)}\,\ee^{-\ii\phii}},
  \label{egeqq}
\end{align}
where $\cV(z,\bar w)$ is the bicomplex function \eqref{vzw} and the equation holds up to neglected $\cO(\ell^4)$ corrections. In the extreme classical limit, the potential expansion \eqref{topop} boils down to $\langle z|PV\!P|w\rangle \sim \cV(z,\bar z)\delta^2(z-w)$, so Eq.~\eqref{egeqq} implies $\Ezo_m = V_{0}(\ell^2m)=V(\gamma_m)$ at leading order. The issue is to find the two remaining unknowns, namely the function $n(f(\phii))$ and the first-order energy correction $\Efo_m$.

To determine these, the crucial step is to evaluate Eq.~\eqref{egeqq} along the equipotential \eqref{epot} labeled by $K = m$, \ie for $z=\sqrt{m f'(\alpha)}\,\ee^{\ii\alpha}$ with $\alpha \in [0,2\pi)$, assuming as before $m\gg1$. Indeed, if Eq.~\eqref{egeqq} holds on a level curve, then it holds for all $z$ by holomorphicity. This is written in more detail in Appendix~\ref{appexp}, where we show that the integrand of Eq.~\eqref{egeqq} has a saddle point at $\phii=\alpha+\cO(1/\sqrt{m})$, eventually resulting in a transport equation for the unknown function $\cn(\theta)$. Here, we skip the computation and analyze separately the real and imaginary parts of the transport equation. We start with the real part, which will allow us to deduce the LLL-projected energy spectrum. The imaginary part is postponed to Sec.~\ref{SubSec:GWF}, where we also display the ensuing nearly Gaussian wave functions.

Let $\Phi(\phii)$ be the phase of $\cn(f(\phii))\equiv\cN(\phii)\,\ee^{\ii\Phi(\phii)}$. Then, the real part of the transport equation [see Eq.~\eqref{rtransport}] yields
\begin{multline}
\label{phiprime}
\Phi'(\phii)
=
\frac{\Efo_{m}}{\Omega_{m}} f'(\phii) 
-\frac{1}{2} \biggl(1+\frac{\Gamma_{m}}{\Omega_{m}}\biggr) \biggl(1+\frac{f''(\phii)^2}{4f'(\phii)^2}\biggr) \\
- \frac{1}{2}
+ \partial_{\phii} \biggl(\frac{f''(\phii)}{8f'(\phii)}\biggr)
+ \frac{1}{2} \frac{\partial_{\phii}[f''(\phii)/2f'(\phii)]}{1+f''(\phii)^2/4f'(\phii)^2},
\end{multline}
where $\Efo_{m}$ is the first-order correction to the energy \eqref{ev} and we introduced the derivatives
\begin{align}
\label{OmegaGamma}
\Omega_{m} \equiv V_{0}'(\ell^2m)>0,
\qquad
\Gamma_{m} \equiv \ell^2m\, V_{0}''(\ell^2m).
\end{align}
In many-body droplets with $N$ electrons, these will respectively measure the Fermi velocity and the curvature of the spectrum at the Fermi surface when $m=N$. Note that all terms in Eq.~\eqref{phiprime} except the factor $1+[f''/2f']^2$ are total derivatives, so the solution is
\begin{multline}
\label{phitap}
\Phi(\phii)
=
\frac{\Efo_{m}}{\Omega_{m}} f(\phii)
-\frac{1}{2} \biggl(1+\frac{\Gamma_{m}}{\Omega_{m}}\biggr) \int_{0}^{\phii}\!\dd\alpha\, \biggl(1+\frac{f''(\alpha)^2}{4f'(\alpha)^2}\biggr) \\
-\frac{\phii}{2}
+\frac{f''(\phii)}{8f'(\phii)}
+\frac{1}{2}\arctan\biggl(\frac{f''(\phii)}{2f'(\phii)}\biggr) + \mathrm{const}.
\end{multline}
This turns out to imply a quantization condition for energy. Indeed, when we initially introduced the function $\cn(\theta)$ in Eq.~\eqref{psim}, we mentioned that it must have a \i{vanishing} winding number along the equipotential, so that all the winding of $\psi_m(\bx)$ is contained in the exponential factor $\ee^{\ii m\theta}$. The phase $\Phi(\phii)$ must therefore be strictly $2\pi$-periodic, \ie $\Phi(2\pi)=\Phi(0)$. Using Eq.~\eqref{phitap}, this fixes the first quantum correction of the energy \eqref{ev}:
\beq
\label{equan}
\frac{\Efo_{m}}{\Omega_{m}}
=
\frac{1}{2}
+
\biggl(1+\frac{\Gamma_{m}}{\Omega_{m}}\biggr)\oint
\frac{\dd\phii}{4\pi}\, \biggl( 1+\frac{f''(\phii)^2}{4f'(\phii)^2} \biggr).
\eeq
The latter generally depends on $m$ through $\Gamma_m$ and $\Omega_{m}$ in Eq.~\eqref{OmegaGamma}. A simplification occurs in `harmonic' setups, where $\Gamma_m=0$ and the right-hand side of Eq.~\eqref{equan} is an $f$-dependent constant, for all $m$ \cite{Note8}. In any case, the full $m^{\text{th}}$ energy \eqref{ev} in the semiclassical limit reads
\begin{multline}
\label{etop}
\!\!\! E_m
\sim
V_{0}\bigl(\ell^2m\bigr) \\
+
\frac{\ell^2}{2}
\biggl[\Omega_m
+
\bigl(\Omega_m+\Gamma_m\bigr)
\oint
\frac{\dd\phii}{2\pi} \biggl(1+\frac{f''(\phii)^2}{4f'(\phii)^2}\biggr)\biggr],
\end{multline}
reproducing the expression announced in Eq.~\eqref{ressum_E_m} with $\lambda(\phii)=2f'(\phii)$, and generalizing the isotropic value obtained for $f'(\phii)=1$ [see Eq.~\eqref{emma}]. The leading-order Bohr-Sommerfeld quantization condition \eqref{oint} is manifestly satisfied, while the first quantum correction can be written in terms of a Maslov-like shift and an integral of the Laplacian, confirming the general result in \cite{Charles:2003b}:
\begin{multline}
\label{emi}
E_m
=
V_{0}\Bigl(\ell^2\Bigl[m+\frac{1}{2}\Bigr]\Bigr) \\
+\frac{\ell^2}{4} \!
\oint\frac{\dd\phii}{2\pi} f'(\phii)\nabla^2V\big|_{r^2=2\ell^2(m+1/2)f'(\phii)}
+\cO(\ell^4).
\end{multline}
(In the language of \cite{Charles:2003b}, our `Maslov-like' term actually stems from an integral of the curvature of $\gamma_m$.)

%---------------------------------------------------------------
\subsection{Gaussian wave functions}
\label{SubSec:GWF}
%---------------------------------------------------------------

As above, write $\cn(f(\phii))=\cN(\phii)\,\ee^{\ii\Phi(\phii)}$ for the unknown function of the WKB ansatz, with a norm $\cN(\phii)=|\cn(f(\phii))|$. Then, the imaginary part of the transport equation [see Eq.~\eqref{itransport}] can be recast into
\beq
\label{eqn}
\frac{\cN'(\phii)}{\cN(\phii)}
=
\frac{1}{4} \partial_{\phii} \log \left[ \frac{1}{f'(\phii)} \left(1+\frac{f''(\phii)^2}{4f'(\phii)^2}\right) \right],
\eeq
which remarkably has the form of an overall logarithmic derivative. The general solution is therefore
\beq
\label{den}
\big|\cn\big(f(\phii)\big)\big|
=
N_0
\left[\frac{1}{f'(\phii)}\left(1+\frac{f''(\phii)^2}{4f'(\phii)^2}\right)\right]^{1/4},
\eeq
where the normalization $N_0$ will soon be fixed. Note the exponent ${1}/{4}$, typical of WKB approximations \cite{BenderOrszag:1999}.

We can now use Eq.~\eqref{den} to evaluate approximate eigenfunctions \eqref{true} near their maximum, \ie close to the equipotential \eqref{epot} with $K=m$. To see this, zoom in on the equipotential by writing
\beq
z
\equiv
\bigl(\sqrt{m}+a\bigr) \sqrt{f'(\alpha)}\,\ee^{\ii\alpha}
\label{zalpp}
\eeq
for $m \gg 1$ and some finite $a$. The integral \eqref{true} then has a unique saddle point at $\phii = \alpha + \delta_1/\sqrt{m} + O(1/m)$, with $\delta_1 = -\ii a \Bigl[ 1 - \ii \tfrac{f''(\alpha)}{2f'(\alpha)} \Bigr]^{-1}$. The saddle-point approximation of the wave function \eqref{true} thus yields
\begin{align}
\psi_m(z,\bar z)
& \sim
\frac{1}{\sqrt{2\pi\ell^2}}\,
\frac{1}{(2\pi m)^{1/4}}\,
\ee^{\ii mf(\alpha) + \ii\Phi(\alpha)} \nonumber \\
& \quad \times
\frac{1}{\sqrt{\sigma(\alpha)}}
\exp \left[ -\frac{f'(\alpha)\,a^2}{1-\ii\frac{f''(\alpha)}{2f'(\alpha)}} \right].
  \label{aspt}
\end{align}
Here, we used Eqs.~\eqref{phitap} and \eqref{den} for the phase and norm of $\cn(f(\alpha))$, fixed the integration constant in Eq.~\eqref{den} to $N_0=\frac{1}{2\pi\ell}(\tfrac{m}{2\pi})^{1/4}$, and introduced the function
\beq
\label{sigma}
\sigma(\phii)^2
\equiv
\frac{1}{f'(\phii)}
\biggl(1+\frac{f''(\phii)^2}{4f'(\phii)^2}\biggr),
\eeq
written earlier in Eq.~\eqref{sigma_lambda} with $\lambda(\phii)=2f'(\phii)$. The normalization was fixed so that the square of the function in Eq.~\eqref{aspt},
\beq
\label{gaussbis}
|\psi_m(z,\bar z)|^2
\sim
\frac{1}{2\pi\ell^2}\,
\frac{\ds\ee^{-2a^2/\sigma^2(\alpha)}}{\sqrt{2\pi m}\,\sigma(\alpha)},
\eeq
is a genuine probability density such that $\int \dd^2\bx\, |\psi|^2 = 1$.

\begin{figure}[t]
\centering
\includegraphics[width=\columnwidth]{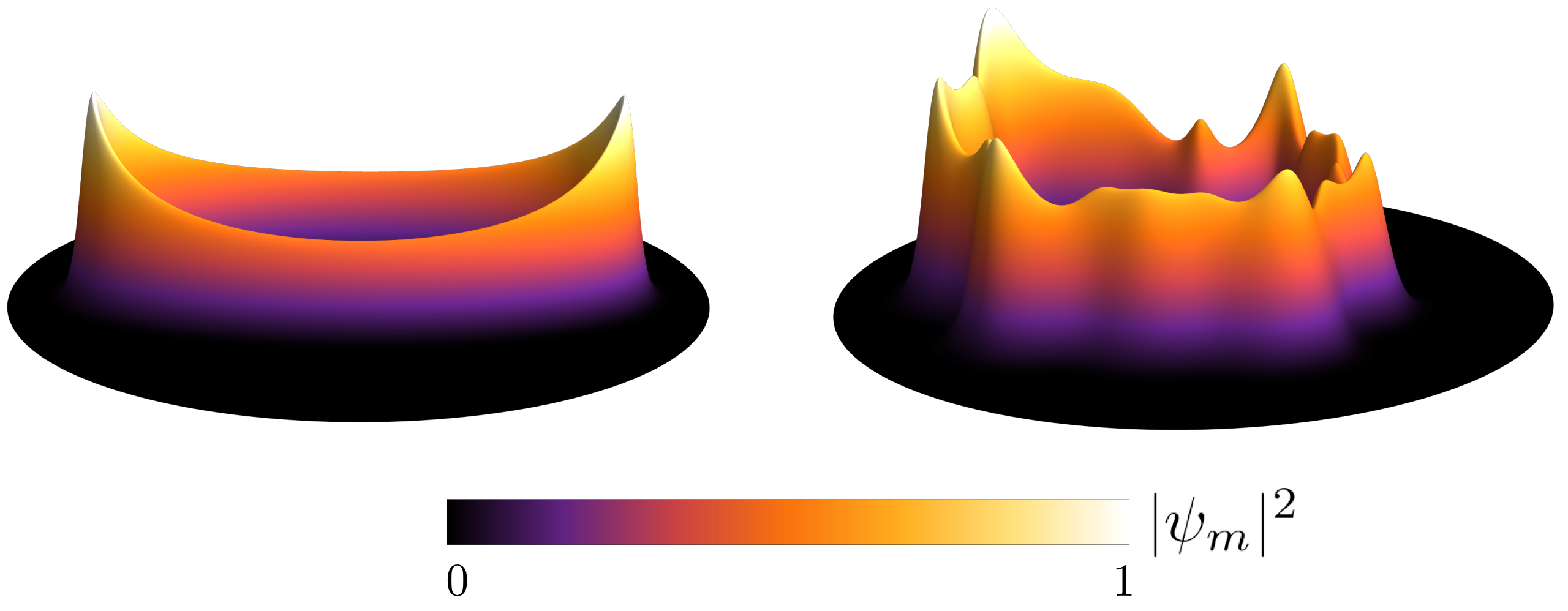}
\caption{The density of a wave function \eqref{aspt} for $m=30$ in an edge-deformed trap \eqref{viviz}. The Gaussian behavior is manifest, as is the angle-dependent `roller coaster' predicted by Eq.~\eqref{gaussbis}, reflecting the changing Euclidean norm \eqref{VELO} of the local velocity of classical guiding centers. (Peaks and troughs of the wave function respectively correspond to minima and maxima of the local velocity.) \i{Left:} Anisotropic harmonic potential given by Eq.~\eqref{viviz} for $f$ of the form \eqref{boost}, with $k=2$ and $\alpha = \cosh(1)$, $\beta = \sinh(1)$. \i{Right:} The same edge-deformed trap as in Figs.~\ref{fiDrop} and \ref{fipott}.}
\label{fiwavefct}
\end{figure}

The wave function \eqref{aspt} coincides with the earlier expression \eqref{e13} upon using $\lambda(\phii)=2f'(\phii)$ and the phase
\beq
\label{Theta_m_x}
\Theta_{m}(\bx)
\equiv
mf(\phii) + \Phi(\phii) - \frac{a^2 f''(\phii)}{2f'(\phii) \sigma(\phii)^2}.
\eeq 
The Gaussian behavior of LLL-projected eigenstates is thus manifest, as anticipated at the end of Sec.~\ref{sseniseg} for the general WKB ansatz \eqref{final}. In that context, we stress again that the exponent $a^2/\sigma^2$ in Eq.~\eqref{gaussbis} is nothing but the squared distance \eqref{edistance} away from the equipotential, while the function $\sigma(\phii)$ defined by Eq.~\eqref{sigma} is essentially the velocity \eqref{VELO} of guiding centers. It is therefore classically expected that the probability of finding an electron at position $\phii$ is proportional to $1/\sigma(\phii)$, which is indeed confirmed by the density \eqref{gaussbis}. As a consequence, the wave function \eqref{aspt} generally behaves as a `roller coaster' whose height follows the local symplectic gradient of the confining potential; this is illustrated in Fig.~\ref{fiwavefct} for two choices of edge-deformed traps \eqref{viviz}. Finally, note that Eq.~\eqref{aspt} generalizes the behavior of isotropic states \eqref{phim} [see Eq.~\eqref{refas}], including the $\cO(1/\sqrt{m})$ contribution that we did not state here but that can be computed by incorporating the next-order term $\delta_2/m$ for the saddle point and repeating the analysis; see Appendix~\ref{Om-1/2corrections} for details.

%---------------------------------------------------------------
\subsection{Comparison with elliptic wave functions}
\label{SubSec:Elliptic_comparison}
%---------------------------------------------------------------

To conclude this section, we now focus on the flower deformations \eqref{boost} and show that the corresponding transport equation is integrable: both the phase \eqref{phitap} and the norm \eqref{den} can be expressed in terms of elementary functions. These results are valuable in themselves since `flowers' are the simplest edge deformations \cite{Note:Fourier}, but also because their special case $k=2$ reproduces known wave functions for anisotropic harmonic traps \cite{DiFrancesco:1994, ForresterJancovici:1996}, providing an important benchmark for our WKB approach.

Consider first the deformation \eqref{boost} with $\alpha = \cosh(\lambda)$ and $\beta = \sinh(\lambda)$ for an arbitrary integer $k$ and a real parameter $\lambda$. When $\lambda>0$, the ensuing potential \eqref{viviz} is steeper in the directions $\phii = 0, \tfrac{2\pi}{k}, \tfrac{4\pi}{k}, \ldots, \tfrac{(2k-2)\pi}{k}$ and lower around the petals at $\phii = \tfrac{\pi}{k}, \tfrac{3\pi}{k}, \tfrac{5\pi}{k}, \ldots, \tfrac{(2k-1)\pi}{k}$; see Fig.~\ref{fidens} for $k=3$. Then, the energy quantization condition \eqref{equan} can be integrated exactly, yielding
\beq
\label{e50}
\frac{\Efo_{m}}{\Omega_{m}}
=
\frac{1}{2}
+
\frac{1}{2}
\bigg(1+\frac{\Gamma_{m}}{\Omega_{m}}\bigg)
\bigg(1+\frac{k^2}{2}\sinh^2(\lambda)\bigg).
\eeq
As for the solution of the transport equation, consisting of the phase \eqref{phitap} and the norm \eqref{den}, it is found to be
\begin{align}
n(\theta)
& =
N_0\,\ee^{-\ii \frac{\Gamma_{m}}{\Omega_{m}} \frac{k}{8} \sinh(2\lambda) \sin(k\theta)}
\ee^{\ii\frac{\theta}{2} \bigl[ 1 - \frac{k}{2} + \bigl(1+\frac{\Gamma_{m}}{\Omega_{m}}\bigr) \bigl(1-\frac{k^2}{4}\bigr) \bigr]} \nonumber \\
& \times
\biggl(\frac{\cosh(\lambda) - \sinh(\lambda)\ee^{\ii k\theta}}{\cosh(\lambda)\ee^{\ii k\theta} - \sinh(\lambda)}\biggr)^{\hspace{-0.7pt}\frac{1}{2k} \bigl[ 1 - \frac{k}{2} + \bigl(1+\frac{\Gamma_{m}}{\Omega_{m}}\bigr) \bigl(1-\frac{k^2}{4}\bigr) \bigr]} \nonumber \\
& \times
\sqrt{\frac{1 + \ii\frac{k}{2}\sinh(2\lambda)\sin(k\theta)}{\cosh(\lambda) - \sinh(\lambda)\ee^{-\ii k\theta}}},
  \label{flosol}
\end{align}
up to an overall constant phase. [Recall that the overall constant is $N_0=\frac{1}{2\pi\ell}(\tfrac{m}{2\pi})^{1/4}$ for normalized wave functions \eqref{aspt}.]

Eq.~\eqref{flosol} depends in a nontrivial way on the potential's derivatives \eqref{OmegaGamma}, with some simplification in the `harmonic' regime $\Gamma_m=0$. Let us therefore apply Eqs.~\eqref{e50} and~\eqref{flosol} to the case of an elliptic harmonic potential, meaning $k = 2$ with constant stiffness $\Omega_m=\Omega>0$ (hence $\Gamma_m=0$). The corresponding edge deformation \eqref{edd} maps the isotropic harmonic potential $V_{0}(r^2/2)=\Omega\,r^2/2$ on its anisotropic cousin,
\beq
\label{ellipsepotential}
V(\bx)
= \Omega \frac{\ee^{2\lambda}x^2+\ee^{-2\lambda}y^2}{2},
\eeq
whose equipotentials are ellipses rather than circles (with their major axis along $y$ for $\lambda>0$). The energy correction \eqref{e50} then becomes $\Efo_m=\Omega\cosh^2(\lambda)$, and the (normalized) solution \eqref{flosol} of the transport equation is
\beq
n(\theta)
=
\frac{1}{2\pi\ell}\Big(\frac{m}{2\pi}\Big)^{1/4}
\sqrt{\cosh(\lambda)+\sinh(\lambda)\ee^{2\ii\theta}}.
\eeq
It is straightforward to use these data to obtain the elliptic version of the normalized Gaussian wave function~\eqref{aspt}:
\begin{align}
\psi_m(z,\bar{z})
& \sim
\frac{1}{\sqrt{2\pi\ell^2}}\,
\frac{1}{(2\pi m)^{1/4}}\,
\frac{\ee^{\ii m\theta}}{\sqrt{\cosh(\lambda) +\sinh(\lambda)\ee^{-2\ii\theta}}} \nonumber \\
& \quad\; \times
\exp\left( - \frac{\ee^{2\ii\theta}-\tanh(\lambda)}{\ee^{2\ii\theta}+\tanh(\lambda)} a^2 \right),
  \label{wfunc_ellipse}
\end{align}
again up to an overall constant phase. Crucially, this coincides with the large-$m$ approximation of the \i{exact} LLL-projected eigenstates of the harmonic potential \eqref{ellipsepotential} \cite{ForresterJancovici:1996}, as can be verified thanks to known asymptotic formulas for Hermite polynomials. This is actually true even at subleading order in $m^{-1/2}$; see Appendix~\ref{Om-1/2corrections:Comp}.

%===============================================================
\section{Many-body observables}
\label{semany}
%===============================================================

This section applies the results of Secs.~\ref{sesdiff} and~\ref{sescale} to entire QH droplets consisting of a large number $N\gg1$ of electrons. Specifically, we exploit our insights on near-Gaussian single-particle wave functions \eqref{aspt} to compute many-body observables and read off the universal shape-dependent effects due to the deformation $f(\phii)$. We first show that the density equals $\frac{1}{2\pi\ell^2}$ in the bulk and drops to zero as an error function at the edge $r_{\text{edge}}=\ell\sqrt{2Nf'(\phii)}$. Second, we turn to the current and show that it is localized as a Gaussian on the edge, to which it is tangent. Third, correlations near the edge are found to display the usual power-law behavior of free fermions, dressed by radial Gaussian factors. This reduces to known expressions in isotropic traps \cite{Cappelli:1992kf}, and to the harmonic results of \cite{DiFrancesco:1994, ForresterJancovici:1996} in the case of flower deformations \eqref{boost} with $k=2$. Finally, the radial behavior of correlations is shown to be consistent with the effective low-energy field theory of edge modes: we derive it microscopically and obtain a chiral CFT in terms of the canonical angle variable on the boundary.

%---------------------------------------------------------------
\subsection{Density}
\label{sedenss}
%---------------------------------------------------------------

\begin{figure}[t]
\centering
\includegraphics[width=.45\textwidth]{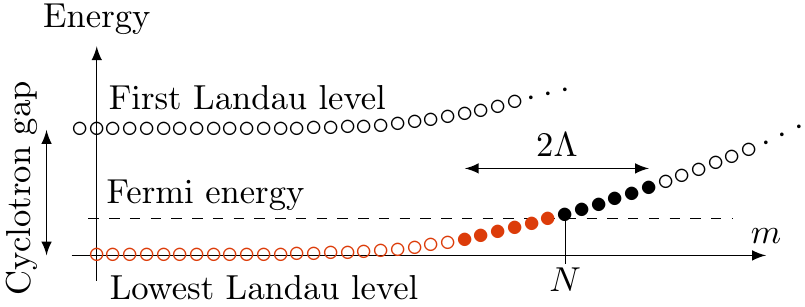}
\caption{The one-body spectrum \eqref{emi}, where the $N$ states that are occupied in the many-body ground state \eqref{OM} are highlighted in red and those that contribute to the low-energy Hamiltonian \eqref{hato} are filled (black for `particles' and red for `holes'). Energies are filled up to a Fermi energy set by a chemical potential $\mu$ such that $E_{N-1}\leq\mu<E_N$. The cutoff $\Lambda$ is large but much smaller than $N$ in the sense that the limit $\Lambda\to\infty$ is taken after taking the thermodynamic limit $N\to\infty$ at fixed $\Lambda$.}
\label{figround}
\end{figure}

Consider a QH droplet of $N\gg1$ noninteracting 2D electrons governed by the Hamiltonian \eqref{h}, with a very strong magnetic field $\bB=\dd\bA$ and a weak edge-deformed potential \eqref{viviz}. The ground state $|\Omega\rangle$ of this many-body system is a Slater determinant of the wave functions $\psi_m$ for occupied states $m=0,1,\ldots,N-1$, where we recall that $m$ is a quantized action variable generalizing angular momentum. This is schematically depicted by red dots in Fig.~\ref{figround}. Explicitly,
\beq
\label{OM}
|\Omega\rangle
=
\prod_{m=0}^{N-1}a^{\dagger}_{0,m}|0\rangle,
\eeq
where $|0\rangle$ is the empty state and $a^{\dagger}_{0,m}$ is a (canonically normalized) Fock space creation operator for the one-particle wave function $\psi_m$ in the LLL. (We will later use $a^{\dagger}_{n,m}$ for creation operators in the $n^{\text{th}}$ Landau level; see Sec.~\ref{Sec:MBO:Edge_modes}.) The same ground state can be obtained by fixing a chemical potential (Fermi energy) $\mu$ and filling all one-body energies $E_m$ such that $E_m\leq\mu$, implying that $N$ is the largest integer such that $E_{N-1}\leq\mu$ \cite{Note9}.

Since each $\psi_m$ yields a single-particle probability density $|\psi_m(\bx)|^2$, the many-body density of the state \eqref{OM} is a sum,
\beq
\label{desum}
\rho(\bx)
=
\sum_{m=0}^{N-1}
|\psi_m(\bx)|^2.
\eeq
While WKB theory does not give access to the form of $\psi_m$ at low $m$, large values of $m$ should be correctly captured by the analysis of Sec.~\ref{sescale}, in which case the one-body density is approximately Gaussian and given by Eq.~\eqref{gaussbis}. We now exploit this Gaussian form to evaluate the many-body density, both in the bulk and close to the edge. (Some technical details are highlighted along the way, as the same method will later allow us to study the many-body current and correlations.)

The key point is that each wave function \eqref{gaussbis} is localized on an equipotential of $V(\bx)$ with area $2\pi\ell^2m$, so the density close to some equipotential $|z| = \mathrm{const} \times \sqrt{f'(\phii)}$ only receives sizeable contributions from wave functions whose quantum number is close to $|z|^2/f'(\phii)$. Accordingly, the bulk density for $1\ll|z|\ll\sqrt{N}$ is obtained by letting the upper summation bound in Eq.~\eqref{desum} go to infinity and writing the approximate density as
\beq
\rho(\bx)
\sim
\frac{1}{2\pi\ell^2}\,
\sum_{m=m_0}^{\infty}
\frac{\ee^{-\frac{2}{\sigma(\phii)^2}\left(\frac{|z|}{\sqrt{f'(\phii)}}-\sqrt{m}\right)^2}}{\sqrt{2\pi m}\,\sigma(\phii)},
\eeq
where the lower summation bound $m_0$ is irrelevant as long as it is large but much smaller than $|z|^2$ and $\sigma(\phii)$ is given by Eq.~\eqref{sigma}. At large $|z|$, the Euler-Maclaurin formula allows us to approximate the sum over $m$ by a (Gaussian) integral over $\sqrt{m}$. This yields the uniform density
\beq
\label{rhobulk}
\rho(\bx)
\sim
\frac{1}{2\pi\ell^2},
\eeq
as expected in the bulk of a QH droplet consisting of a single fully filled LLL \cite{Note10}.

\begin{figure*}[t]
\centering
\includegraphics[width=\textwidth]{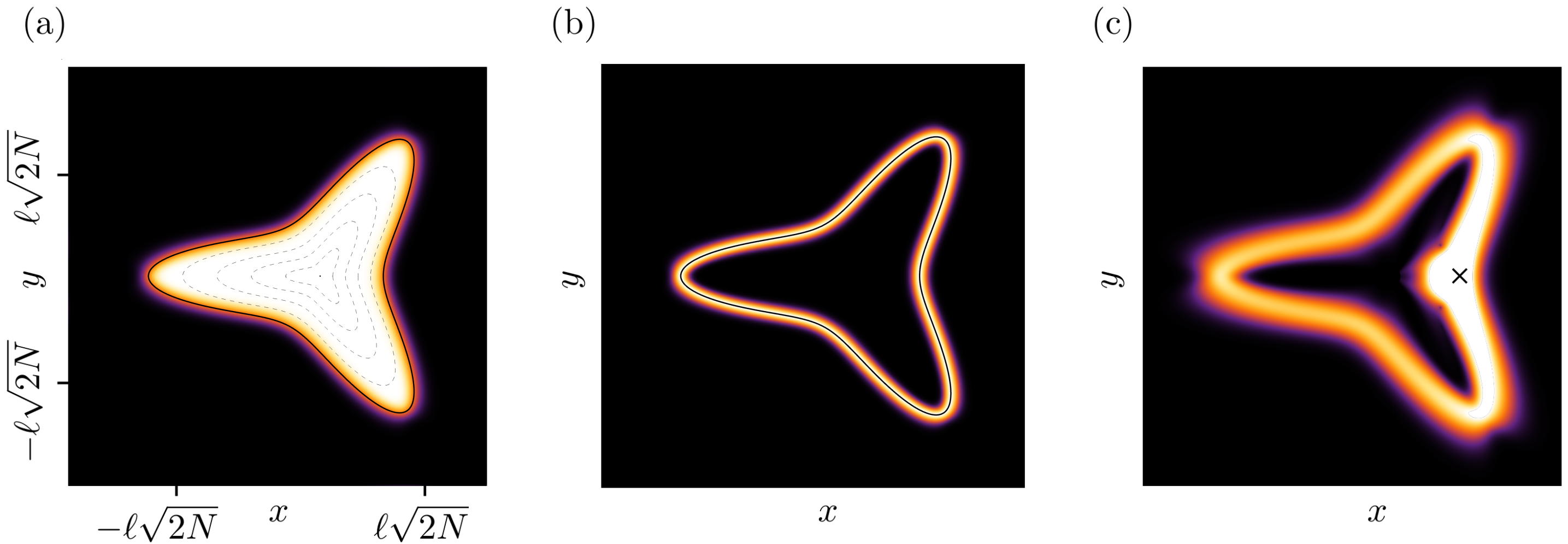}
\caption{Intensity plots of: (a) The density \eqref{rhoo} for $N=100$ electrons and a flower deformation \eqref{boost} of order $k=3$ with $\alpha = \cosh(1)$ and $\beta = \sinh(1)$. The constancy of density in the bulk and its sharp decay at the boundary are manifest. Note also the depletion of electrons between petals, due to the steepness of the potential at $\phii=0,\tfrac{2\pi}{3},\tfrac{4\pi}{3}$, to be contrasted with the deeper potential wells near the petals at $\phii=\tfrac{\pi}{3},\pi,\tfrac{5\pi}{3}$. (b) The current's norm \eqref{normj} for the same droplet. The localization on the edge equipotential (black curve) is clearly visible, as is the width of the Gaussian jump. (c) The norm of the correlation function \eqref{cofoBis} for the same droplet, seen as a function of $\bx_2$ when $\bx_1=(\ell\sqrt{2Nf'(0)},0)$ (black cross) is fixed close to the edge; its behavior for well-separated points is given by the long-range correlator \eqref{coco}. The color coding is the same as in Fig.~\ref{fipott}.}
\label{fidens}
\end{figure*}

An analogous argument can be carried out close to the droplet's edge, with one key difference: the upper summation bound in Eq.~\eqref{desum} is now crucial. Thus, letting $|z|=\bigl(\sqrt{N}+a\bigr)f'(\phii)$ with finite $a$ in the large-$N$ limit and using once more the approximate Gaussian form \eqref{gaussbis}, the density \eqref{desum} near the edge behaves as
\beq
\label{edgerho}
\rho(\bx)
\sim
\frac{1}{2\pi\ell^2}\,
\sum_{k=1}^{\infty}
\frac{\ee^{-\frac{2}{\sigma(\phii)^2}\left(a+\frac{k}{2\sqrt{N}}\right)^2}}{\sqrt{2\pi N}\,\sigma(\phii)},
\eeq
where we changed variables as $m \equiv N-k$ with $k = \cO(\sqrt{N})$ at large $N$ and only kept track of leading-order terms. For $N \gg 1$, the sum over $k$ can once more be converted into an integral, now over $k/2\sqrt{N}$. This yields the asymptotic behavior
\beq
\label{rhoo}
\rho(r,\phii)
\sim
\frac{1}{4\pi\ell^2}
\erfc\left(
\frac{1}{\sigma(\phii)}
\frac{r - \ell\sqrt{2Nf'(\phii)}}{\ell\sqrt{f'(\phii)}}
\right),
\eeq
where $\erfc$ denotes the complementary error function and the width \eqref{sigma} is inherited from that of our LLL wave functions. This explicit result was announced in Eq.~\eqref{ressum_rho_edge} with $\lambda(\phii)=2f'(\phii)$. It confirms that the density is roughly constant and given by Eq.~\eqref{rhobulk} in the bulk, then drops to zero within a distance of the order of the magnetic length \eqref{ell} around the edge at $r=\ell\sqrt{2Nf'(\phii)}$; see Figs.~\ref{fipott}(a) and \ref{fidens}(a).

We stress that, in contrast to wave functions, the density \eqref{rhoo} only depends on the potential near the edge of the droplet: bulk deformations of the potential do not affect the quantized bulk density \eqref{rhobulk} in the limit of strong magnetic fields. In this sense, Eq.~\eqref{rhoo} is a universal formula for the density of any QH droplet of LLL states whose edge traces an equipotential of the form $r^2 = 2\ell^2Nf'(\phii)$; such an error function behavior is indeed known to hold in considerable generality \cite{ZelditchZhou}. It would be instructive to probe this local density in experiments, using either real samples \cite{Hashimoto2008, Hashimoto} or quantum simulators \cite{Goldman1, Goldman2, Fletcher, Mukherjee:2021jjl}.

Note that the leading-order result \eqref{rhoo} receives a number of subleading corrections that can be systematically computed in our formalism; these are omitted here for brevity, but the $\cO(1/\sqrt{N})$ correction is evaluated in Appendix~\ref{Om-1/2corrections:MB}. A related comment is that Eqs.~\eqref{rhobulk} and \eqref{rhoo} are only valid at extremely strong magnetic fields, which stems from the simplification provided by the LLL projection. The actual density profile, both in the bulk and near the edge, depends on the gradient of the potential. For instance, anharmonic traps \cite{Nardin_carusotto_2020} give rise to an excess charge density at the edge, but this involves higher Landau levels that are beyond our scope. Interactions are similarly absent here, so electrostatic backreaction and edge reconstruction \cite{Chklovskii} do not appear in our approach. In this respect, the application of Eq.~\eqref{rhoo} to real condensed-matter samples is subtle; quantum-engineered systems may provide a better platform to observe such detailed local effects.

%---------------------------------------------------------------
\subsection{Current}
\label{securr}
%---------------------------------------------------------------

The current of a droplet of $N\gg1$ electrons can similarly be expressed as a sum over single-particle currents. To this end, recall that the gauge-invariant one-body probability current of a charged wave function $\psi$ with mass $M$ is a one-form $\hbar\,\bj/M$ given by
\beq
\label{john}
\bj
=
\frac{1}{2\ii}
\left( \psi^*\dd\psi - \psi\dd\psi^* - 2\ii\frac{q}{\hbar}\bA|\psi|^2 \right),
\eeq
where the first term is only sensitive to the gradient of the phase of $\psi$ and $\bA = (Br^2/2)\,\dd\phii = (\hbar/q)|z|^2\,\dd\phii$ in symmetric gauge. The many-body current of the ground state \eqref{OM} is thus
\beq
\label{johnson}
\bJ
=
\sum_{m=0}^{N-1}\bj_m,
\eeq
where $\bj_m$ is the single-particle current \eqref{john} of each occupied one-particle state $\psi_m$. 

As before, the WKB approximation does not give access to wave functions for small $m$, but this is unimportant close to the edge. In that regime, we have already gathered all the ingredients needed to evaluate the currents \eqref{john} up to small quantum corrections: the one-body density is given by Eq.~\eqref{gaussbis}, while the derivative of the phase is obtained from Eq.~\eqref{Theta_m_x} and the real part \eqref{phiprime} of the transport equation. In practice, the WKB phase $\Phi$ turns out to be negligible at leading order, and the only relevant parts of the phase are those explicitly visible in Eq.~\eqref{aspt}: the (fast) phase $\ee^{\ii m f(\phii)}$ together with the contribution from $\bA = \bigl(\hbar/q\bigr)|z|^2\, \dd\phii$ eventually gives rise to the leading angular component of the current, while the (slow) phase $\ee^{-\ii [ f''(\phii)/2f'(\phii)] a^2/\sigma^2}$ yields its radial component that is nonzero whenever $f''(\phii) \neq 0$.

Starting from these facts, it is straightforward to adapt the method of Sec.~\ref{sedenss} to the many-body current \eqref{johnson}. Writing $|z| = \bigl(\sqrt{N}+a\bigr) \sqrt{f'(\phii)}$, the sum over $m \equiv N-k$ becomes an integral over $k/2\sqrt{N}=\cO(1)$ and yields the leading-order result \eqref{ressum_J_edge} with $\lambda(\phii)=2f'(\phii)$:
\beq
\label{joo}
\bJ(r,\phii)
\sim
-\frac{\ds\ee^{-2a^2/\sigma(\phii)^2}}{(2\pi \ell^2)^{3/2}\sigma(\phii)}
\frac{\ell\sqrt{2Nf'(\phii)}\,\dd\phii+\tfrac{f''(\phii)}{2f'(\phii)}\,\dd r}{\sqrt{2f'(\phii)}},
\eeq
where $a=\bigl(r-\ell\sqrt{2N f'(\phii)}\bigr)\big/\ell\sqrt{2f'(\phii)}$ and $\sigma(\phii)$ is given by Eq.~\eqref{sigma}. Both components in Eq.~\eqref{joo} receive subleading corrections that are omitted here. In particular, there is an $\cO(1)$ term in $J_{\phii}$ that is nonzero on the edge, even in the isotropic case $f'=1$. Evaluating that term requires the $\cO(1/\sqrt{m})$ correction that was neglected in Eq.~\eqref{gaussbis}; see Appendix~\ref{Om-1/2corrections} for the computation of this correction and its contribution to the current.

Using the metric $\dd s^2=\dd r^2+r^2\dd\phii^2$, one can verify that the one-form $\ell\sqrt{2N\,f'}\,\dd\phii + ({f''}/{2f'})\,\dd r$ in Eq.~\eqref{joo} is the dual of a vector tangent to the equipotential at the droplet's edge \cite{Note11}. Moreover, the norm squared
\beq
\label{normj}
\big\|\bJ(r,\phii)\big\|^2
\sim
\frac{1}{2(2\pi\ell^2)^3}
\exp\left[-\frac{2\bigl(r - \ell\sqrt{2N f'(\phii)}\bigr)^2}{\ell^2 \sigma(\phii)^2 f'(\phii)}\right]
\eeq
shows that the current has a constant maximum along the edge, with a constant nonzero width owing to Eq.~\eqref{edistance}; see Figs.~\ref{fipott}(b) and \ref{fidens}(b).

Similarly to the density, it is important to remember that the LLL projection misses some important physics. Indeed, the actual bulk current is the symplectic gradient of the confining potential multiplied by the Hall conductance \cite{Geller1, Geller2, Champel1, Champel2}. No such effect occurs in Eq.~\eqref{joo} because it requires higher Landau levels, which are beyond our scope.

%---------------------------------------------------------------
\subsection{Correlations}
\label{secor}
%---------------------------------------------------------------

The methods that we applied to density and current can also be used to compute electronic correlations near the edge, for which much less is known. Indeed, consider as before an anisotropic droplet whose occupied one-body states have quantum numbers $m=0,1,\ldots,N-1$. Then, the correlation function between the points $\bx_1$ and $\bx_2$ is
\beq
C(\bx_1,\bx_2)
=
\sum_{m=0}^{N-1}
\psi_m^*(\bx_1)\,\psi_m(\bx_2),
\label{cofoBis}
\eeq
which reduces to the density \eqref{desum} when $\bx_1=\bx_2$. As before, we rename $m\equiv N-k$ and let the complex coordinates $z,w$ corresponding to $\bx_1,\bx_2$ be such that
\beq
\begin{aligned}
z
& =
\bigl(\sqrt{N}+a\bigr)\sqrt{f'(\phii_{1\vphantom{2}})}\,\ee^{\ii\phii_1}, \\
w
& =
\bigl(\sqrt{N}+b\bigr)\sqrt{f'(\phii_2)}\,\ee^{\ii\phii_2},
\end{aligned}
\eeq
where $a,b$ are finite at large $N$ and $\phii_1,\phii_2$ are the polar angles of $\bx_1,\bx_2$. One can then plug the Gaussian wave functions \eqref{aspt} into Eq.~\eqref{cofoBis}, this time assuming $k$ finite, and perform the sum over $k$. The gradient expansion of the potential implies that the ratio $\Gamma_{m}/\Omega_{m}\sim\Gamma_N/\Omega_N+\cO(\ell^2)$ is nearly constant in this regime, so Eq.~\eqref{cofoBis} becomes a geometric sum over $k$ that reproduces the result stated in Eq.~\eqref{ressum_C_edge} with $\lambda(\phii)=2f'(\phii)$:
\begin{align}
C(\bx_1,\bx_2)
& \sim
\frac{\ee^{\ii\Theta_{N}(\bx_1,\bx_2)}}{(2\pi)^{3/2}\ell^2\sqrt{N}}
\frac{1}{\sqrt{\sigma(\phii_1)\sigma(\phii_2)}} \nonumber \\
& \quad \times
\frac{\ii \exp \Bigl( -\frac{a^2}{\sigma(\phii_1)^2}-\frac{b^2}{\sigma(\phii_2)^2} \Bigr)}{2\sin\big([f(\phii_1)-f(\phii_2)]/2\big)},
  \label{coco}
\end{align}
where $\sigma(\phii)$ was defined in Eq.~\eqref{sigma}. The overall phase $\Theta_{N}(\bx_1,\bx_2) = \Theta_{N}(\bx_2) - \Theta_{N}(\bx_1)-[f(\phii_2)-f(\phii_1)]/2$, given by Eq.~\eqref{Theta_m_x}, involves the WKB phase \eqref{phitap}.

Several features of Eq.~\eqref{coco} are worth emphasizing. First, note the striking appearance of long-range correlations localized at the edge by a Gaussian envelope. Their power-law decay along the boundary is a static diagnostic of the presence of edge modes \cite{Cappelli:1992kf, Wen, Estienne:2021qqe} and matches the standard CFT propagator written in the angle variable $\theta = f(\phii)$, as will be discussed in Sec.~\ref{Sec:MBO:Edge_modes}. In fact, this behavior agrees with the classical picture in which edge modes propagate at constant speed in the $\theta$ coordinate \cite{Note13}. A second key aspect is the lack of translation invariance in $\phii$ along the edge, caused not only by the argument $f(\phii_1) - f(\phii_2) = \int_{\phii_2}^{\phii_1} \dd \phii\, f'(\phii)$ but also by the factors $\sigma(\phii_1)$ and $\sigma(\phii_2)$. Interestingly, the product $\sigma(\phii_1)^{-1/2}\sigma(\phii_2)^{-1/2}$ resembles prefactors picked up by primary fields in CFT under local conformal transformations.

It is tempting to compare our result with the edge correlations of a 2D Coulomb gas of the same shape as the droplet. For harmonic potentials, the two problems are indeed identical, and Eq.~\eqref{coco} coincides with the plasma result of \cite{ForresterJancovici:1996} upon using the map \eqref{boost} with $k=2$, $\alpha = \cosh(\lambda)$, and $\beta = \sinh(\lambda)$. But there is no such coincidence for arbitrary edge deformations: our formula \eqref{coco} generally differs from the plasma prediction of \cite{Jancovici:1995}. [This can be verified \eg with flower deformations \eqref{boost} of order $k\geq3$.] The reason for this mismatch is that Coulomb-gas edge correlations are obtained by solving an electrostatic problem that only depends on the shape of the droplet \cite{Jancovici:1995}, whereas QH edge correlations are sensitive not only to a droplet's shape, but also to the local edge velocity, which is controlled by the gradient of the potential. These ideas will be further explored elsewhere.

Finally, it is a simple matter to include time dependence in the correlator \eqref{coco}. Indeed, the occupied one-particle states in Eq.~\eqref{cofoBis} have definite energies $E_m$ given by Eq.~\eqref{emi} at large $m$. This spectrum is approximately linear close to the Fermi energy: changing variables according to $m = N+k$ with $k$ finite at large $N$, one has
\beq
\label{enek}
E_{N+k} - E_N
\sim
\hbar \omega_{\text{F}} k
\eeq
with $\omega_{\text{F}} \equiv \ell^2\Omega_N/\hbar$ the angular Fermi velocity given by the potential's derivative in Eq.~\eqref{OmegaGamma} at $m=N$. In the linear regime \eqref{enek}, one can repeat the asymptotic computation of correlations to find once more an expression of the form \eqref{coco}, now with a time-dependent overall phase and a time-dependent denominator $2\sin\big([f(\phii_1)-f(\phii_2)-\omega_{\text{F}}(t_1-t_2)]/2\big)$. This exhibits the standard ballistic propagation of correlations in a CFT, which we confirm below from the low-energy dynamics of our droplet.

%---------------------------------------------------------------
\subsection{Edge modes}
\label{Sec:MBO:Edge_modes}
%---------------------------------------------------------------

The effective low-energy description of anisotropic QH droplets can be derived similarly to the isotropic case \cite{Cappelli:1992kf} inspired by Luttinger-liquid theory \cite{Haldane}. This has the advantage of circumventing topological field theory, at the cost of failing to apply for fractional QH states \cite{Bahcall:1991an, Frohlich, Wen:1990qp, Wen:1990se, Wen}. We now provide such a first-principles calculation, eventually concluding that edge modes span a free chiral CFT expressed in terms of the canonical angle coordinate $\theta=f(\phii)$ along the boundary. Aside from its intrinsic interest, this provides an independent check of the validity of the correlator \eqref{coco}.

Our starting point is the one-body Hamiltonian $H_{\text{one-body}} - \mu$ given by Eq.~\eqref{h} with the chemical potential $\mu$ introduced in Sec.~\ref{sedenss}. The corresponding second-quantized Hamiltonian in the fermionic Fock space is
\beq
\label{hham}
\cH
=
\sum_{n,m\geq0}(E_{n,m}-\mu)a^{\dagger}_{n,m}a\pdag_{n,m},
\eeq
where each $E_{n,m}$ is a one-particle eigenvalue of the operator in Eq.~\eqref{h} labeled by the Landau-level index $n\in\mathbb{N}$ and the `action variable' quantum number $m\in\mathbb{N}$ within each level. (Thus, the eigenvalues found in Sec.~\ref{sescale} are really $E_m\equiv E_{0,m}$.) As for the Fock space operator $a^{(\dagger)}_{n,m}$, it annihilates (creates) the corresponding eigenstate, with standard anticommutation relations $\{a\pdag_{n,m\ppr},a^{\dagger}_{n',m'}\}=\delta_{n,n'}\delta_{m,m'}$. The exact energy spectrum is unknown, but this is not an issue since low-energy excitations all belong to the LLL, with an approximately linear dispersion \eqref{enek} near the Fermi momentum; see Fig.~\ref{figround}. As a consequence, the low-energy approximation of the many-body Hamiltonian \eqref{hham} can be written as
\beq
\label{hato}
\cH
\sim
\sum_{p\in[-\Lambda,\Lambda]}
\hbar\omega_{\text{F}}\,p
\wick{ a^{\dagger}_{p}a\pdag_{p} },
\eeq
where $\Lambda$ is some cutoff on the half-integer label $p$ with $a^{(\dagger)}_{p} \equiv a^{(\dagger)}_{0, N + p-1/2}$ and the right-hand side is normal ordered with respect to the ground state \eqref{OM}---this merely subtracts a constant such that the Hamiltonian has a well-defined $\Lambda\to\infty$ limit \cite{Note12}.

Up to the cutoff $\Lambda$, the Hamiltonian \eqref{hato} is that of a 1D chiral CFT. However, one should keep in mind that the operators $a^{\dagger}_{p}$ create 2D states. Showing the emergence of a truly 1D effective theory relies on the fact that the relevant wave functions \eqref{aspt} are Gaussians localized on the edge. Indeed, provided $\Lambda$ is kept finite while taking the thermodynamic limit $N\to\infty$, one can write the creation operators in Eq.~\eqref{hato} as Fourier modes
\beq
\label{a_Psi_def}
a^{\dagger}_p
=
\oint\frac{f'(\phii)\dd\phii}{\sqrt{2\pi}}\,
\ee^{\ii pf(\phii)}
\Psi^{\dagger}(f(\phii))
\eeq
of an `edge field' $\Psi^{\dagger}(f(\phii))$ that is independent of $p$. This 1D field is a radial integral of 2D creation operators $c^{\dagger}(\bx)$ weighted by the $N^{\text{th}}$ wave function $\psi_N(\bx)$ in Eq.~\eqref{aspt}:
\beq
\label{efild}
\Psi^{\dagger}(f(\phii))
\equiv
\frac{\sqrt{2\pi}}{f'(\phii)}\ee^{-\ii f(\phii)/2}
\int_0^{\infty} r\dd r\,c^{\dagger}(\bx)\psi_N(\bx).
\eeq
We emphasize that the appearance of a single label $N$ in this definition crucially stems from the restriction to quantum numbers that differ from $N$ by a finite amount in the thermodynamic limit. It is then clear that the operator $\Psi^{\dagger}(\theta)$ creates an electron at the position $\theta=f(\phii)$ on the edge. Furthermore, the normalization of the 1D field \eqref{efild} is canonical in angle variables: using the standard anticommutator $\{c(\bx_1),c^{\dagger}(\bx_2)\}=\delta^{(2)}(\bx_1-\bx_2)$, one similarly finds $\{\Psi(f(\phii_1)),\Psi^{\dagger}(f(\phii_2))\}=\delta(f(\phii_1)-f(\phii_2))$ in terms of the Dirac delta function on the circle. This is consistent with the canonical anticommutator of the operators \eqref{a_Psi_def}. Also note that the half-integer labels in Eq.~\eqref{hato} imply antiperiodic (Neveu-Schwarz) boundary conditions in $\phii$ [or $\theta=f(\phii)$] for the edge field \eqref{efild}.

The derivation of the low-energy effective field theory is now essentially done. Indeed, removing the cutoff by taking $\Lambda \to \infty$ in Eq.~\eqref{hato} yields
$\cH_{\text{eff}}
=
\sum_{p}
\hbar\omega_{\text{F}} p
\wick{ a^{\dagger}_{p}a\pdag_{p} }$, where the sum is over all half-integer `momenta' $p \in \mathbb{Z} + 1/2$ and the angular Fermi velocity is $\omega_{\text{F}}=\ell^2\Omega_{N}/\hbar$ with $\Omega_{N}$ given by Eq.~\eqref{OmegaGamma}. This can be recast as
\beq
\label{efeham}
\cH_{\text{eff}}
=
\hbar \oint \dd\theta\, \wick{ \Psi^{\dagger}(\theta)\,(-\ii \omega_{\text{F}}\der_{\theta})\Psi(\theta) }
\eeq
in terms of the 1D fields $\Psi^{(\dagger)}(\theta)$. The canonical normalization of \eqref{efild} then implies the presence of the usual term $\ii\Psi^{\dagger}\der_t\Psi$ in the fermionic action functional of edge modes, which reads
\beq
\label{lactose}
S[\Psi,\Psi^{\dagger}]
=
\hbar
\int\dd t \dd\theta\,\ii\Psi^{\dagger}(\theta)
\big(\der_t+\omega_{\text{F}}\der_{\theta}\big)\Psi(\theta).
\eeq
This is manifestly a local 1D free chiral CFT in terms of the angle variable $\theta=f(\phii)$. By bosonization, the corresponding edge density waves (magnetoplasmons) are similarly described by a free bosonic chiral CFT.

We stress that the simplification leading from highly anisotropic, inhomogeneous wave functions \eqref{aspt} to the homogeneous field theory \eqref{lactose} stems from delicate cancellations between radial and angular dependencies in the integral \eqref{efild}. This generalizes the known theory of edge modes in highly symmetric geometries to the anisotropic setups studied here. The low-energy effective theory \eqref{lactose} is thus universal: for any trapping potential, edge modes are described by a chiral fermionic CFT expressed in terms of the canonical angle coordinate of the trap at the boundary. One could have guessed this from the dynamics of electronic guiding centers induced by the potential $V$ in the noncommutative plane \eqref{pxp} \cite{Note13}. In the present case, the angle coordinate is $\theta=f(\phii)$; more general cases involve more complicated action-angle variables.

Of course, $\theta$ generally has nothing to do with other obvious position coordinates, such as the polar angle $\phii$ or the arc length
\beq
s(\phii)
=
\ell\sqrt{2N}
\int_0^{\phii}\dd\alpha\,\sqrt{f'(\alpha)+\frac{f''(\alpha)^2}{4f'(\alpha)}}.
\eeq
Any such `wrong' coordinate makes the apparent Fermi velocity of edge modes position dependent. For example, the Euclidean norm \eqref{VELO} of the velocity of edge modes varies along the boundary and is, in fact, proportional to the function $\sigma(\phii)$ defined in Eq.~\eqref{sigma}. This is reminiscent of inhomogeneous CFTs, whose light cones are curved owing to the presence of a nonzero spacetime curvature \cite{AllegraEtAl2016, DSVC:2017, DSC:2017curvedlightcones, GLM:2018, LangmannMoosavi:2019, Moosavi:2021iCFT, GMS:2022, Moosavi:2023DBdG-iQLs}. However, one should keep in mind that our edge modes sense a \i{flat} metric $\omega_{\text{F}}^2\dd t^2-\dd\theta^2=\omega_{\text{F}}^2\dd t^2-f'(\phii)^2\dd\phii^2$, whose light cones are straight lines in terms of the canonical angle coordinate $\theta=f(\phii)$.

This observation is also consistent with the seemingly complicated correlator \eqref{coco}. Indeed, one can start from the definition \eqref{efild} to write the 1D correlation function $\langle \Psi^{\dagger}(\theta_1)\Psi(\theta_2) \rangle$ as a double radial integral of the 2D quantity $\langle c^{\dagger}(\bx_1)c(\bx_2) \rangle$. The asymptotic relation \eqref{coco} then yields numerous simplifications, eventually giving
\beq
\label{intcor}
\langle\Psi^{\dagger}(\theta_1)\Psi(\theta_2)\rangle
=
\frac{1}{2\pi}\,
\frac{\ii}{2\sin\big([\theta_1-\theta_2]/2\big)}.
\eeq
The same result would have been obtained directly from the low-energy action \eqref{lactose}: it is a correlation function of free gapless fermions written in terms of the angle coordinates $\theta_1=f(\phii_1)$ and $\theta_2=f(\phii_2)$. As a bonus, time-dependent correlations automatically satisfy the behavior $\propto\sin\big([\theta_1-\theta_2-\omega_{\text{F}}(t_1-t_2)]/2\big)^{-1}$ stated at the end of Sec.~\ref{secor}.

In conclusion, it is worth noting that the CFT action \eqref{lactose} is only the leading part of the full effective action of edge modes. The latter actually contains many subleading terms---irrelevant corrections that vanish in the thermodynamic limit and typically break conformal invariance. For example, the Taylor expansion \eqref{enek} of energies near the Fermi momentum can be pushed further to include quantum corrections that we have so far neglected. Their leading part can be consistently computed from Eq.~\eqref{etop}, yielding
\beq
\label{enek_nlin}
E_{N+k} - E_N
=
\hbar \left(\tilde\omega_{\text{F}} k + \frac{\ell^4}{2\hbar} \frac{\Gamma_N}{N \ell^2} k^2 + O\bigl(\ell^6\bigr) \right),
\eeq
which involves the spectrum's curvature $\Gamma_N$ in Eq.~\eqref{OmegaGamma} as well as the corrected Fermi velocity
\beq
\label{omega_F_sd}
\tilde\omega_{\text{F}}
=
\omega_{\text{F}}
  + \frac{\ell^4}{2\hbar}
    \left[
      \frac{\Gamma_{N}}{N\ell^2}
      + \frac{2\Gamma_{N} + \Delta_N}{N\ell^2}
        \oint \frac{\dd\phii}{2\pi}\, f'(\phii) \sigma(\phii)^2
    \right].
\eeq
Here, $\omega_{\text{F}}\equiv\ell^2\Omega_N/\hbar$ was defined below Eq.~\eqref{enek} in terms of the potential's derivative in Eq.~\eqref{OmegaGamma}, and $\Delta_N \equiv \ell^4 N^2\, V_{0}'''(\ell^2 N)$. Thus, edge modes have a weak dispersion \eqref{enek_nlin} governed by the curvature of the potential---a well-known effect whose bosonization gives rise to nonlinear dynamics for edge density waves \cite{Wiegmann, Nardin_carusotto_2020, Nardin_carusotto_2023}. The velocity $\omega_{\text{F}}$ itself also receives corrections in Eq.~\eqref{omega_F_sd}; these depend on the entire shape of the droplet through the integral of $f'(\phii) \sigma(\phii)^2$. In principle, since the propagation velocity on the edge is measurable in QH experiments \cite{AshooriEtAl:1992, KataokaEtAl:2016, HashisakaEtAl:2017} and also affects transport \cite{Maclure:2009}, this opens up the possibility to probe the anisotropy \eg in cold-atom realizations \cite{TaiEtAl:2017, LeonardEtAl:2023, BinantiEtAl:2023}, where the smaller number of particles enhances subleading effects. It is somewhat remarkable that our leading WKB wave functions \eqref{aspt} suffice to predict such detailed corrections. In practice, realistic QH systems involve numerous other sources of modifications to the leading effective action \eqref{lactose}, typically stemming from higher Landau levels and/or interactions; no such complications occur here.

%===============================================================
\section{Microwave absorption}
\label{ARMA}
%===============================================================

Our study so far focused on local properties of QH droplets in space---their wave functions, density, etc. While the measurement of such local quantities may soon be within reach in quantum simulators, it is generally much trickier in genuine solid-state systems where one's control over the confining potential is limited. The present section is therefore devoted to an experimentally realistic, global probe of anisotropy that requires \i{no local imaging}. Namely, we consider the microwave absorption spectrum of anisotropic QH droplets \cite{Cano} and show that it consists of a characteristic series of peaks whose magnitude depends on the waves' polarization. This is expected to soon be observable in setups that build upon the experimental work \cite{Mahoney}. In what follows, we first review the basics along with known isotropic results, then turn to anisotropic droplets, finally concluding with nonuniform generalizations and a discussion of subleading effects.

%---------------------------------------------------------------
\subsection{Transition rates and isotropic benchmark}
%---------------------------------------------------------------

Consider a QH droplet of mesoscopic size, with a confining potential such that the angular Fermi velocity $\omega_{\text{F}}$ falls in the microwave range \cite{KamataEtAl:2010, Note:Magnetoplasmon_velocity}. The droplet, prepared in its ground state \eqref{OM}, is placed next to a coplanar microwave transmission line that subjects it to electromagnetic pulses with a frequency $\omega$ close to $\omega_{\text{F}}$. Suddenly switching on the radiation at time $t=0$ excites edge density waves, leading to a frequency-dependent absorption rate $\Gamma(\omega)$ of electromagnetic waves by the droplet. Our goal is to compute this rate.

As a general starting point, let a quantum system be prepared in a given energy eigenstate $|\psi_m\rangle$, and subject it to a time-dependent perturbation $W\cos(\omega t)$ starting at time $t=0$, where $W$ is some Hermitian operator. Standard perturbation theory then predicts that the transition rate from $|\psi_m\rangle$ to some other eigenstate $|\psi_n\rangle$ is
\beq
\label{gone}
\Gamma_{m\to n}
\sim
\frac{1}{2\hbar}
\big|\langle\psi_m|W|\psi_n\rangle\big|^2\,
\delta\big(\hbar\omega-|E_m-E_n|\big)
\eeq
up to subleading corrections involving higher powers of $W$, where $E_m$ and $E_n$ are the respective energies of $|\psi_m\rangle$ and $|\psi_n\rangle$. In particular, Eq.~\eqref{gone} applies to electronic orbital transitions in a QH sample subjected to electromagnetic waves \cite{Tran, TranBis}. Here, we shall mostly focus on linearly polarized, uniform perturbations, in which case $W = qE [x\cos(\alpha) + y\sin(\alpha)]$ in terms of the electric field's amplitude $E$ and the polarization angle $\alpha$. The ensuing transition rates \eqref{gone} can be evaluated thanks to the knowledge of the relevant eigenfunctions. The corresponding many-body absorption rate, for a QH droplet consisting of many states $|\psi_0\rangle, \ldots, |\psi_{N-1}\rangle$, will be a sum of those one-body rates \eqref{gone} that are permitted by the Pauli exclusion principle given the ground state \eqref{OM}.

Let us derive such an absorption rate in the simplest case of an \i{isotropic} QH droplet. Then, there is no loss of generality in assuming polarization along the $x$ axis, and the relevant wave functions \eqref{phim} behave near their maximum as predicted by Eq.~\eqref{aspt} with $f(\phii)=\phii$ and $\Phi=0$. It follows that the matrix elements needed in Eq.~\eqref{gone} satisfy the selection rule $\langle\psi_m|x|\psi_n\rangle\sim\ell\sqrt{m/2}(\delta_{m,n-1}+\delta_{m,n+1})$ at large quantum numbers $m,n$ \cite{Note:Largem}. Note that the Kronecker deltas on the right-hand side would occur for any pair of states with definite angular momenta $m,n$, while the coefficient in front is specific to the LLL. Owing to the selection rule, the only transition allowed by the Pauli exclusion principle for the many-body ground state \eqref{OM} is the one where the state $|\psi_{N-1}\rangle$ jumps to the state $|\psi_N\rangle$; any other transition either is forbidden or occurs at higher orders in the perturbation. Eq.~\eqref{gone} then predicts that the droplet's absorption rate at frequency $\omega$ is given by \cite{Cano}
\beq
\label{giso}
\frac{\Gamma(\omega)}{2\pi N\ell^2}
\sim
\frac{q^2E^2}{8\pi\hbar^2}\,
\delta\big(\omega-\omega_{\text{F}}\big)
\eeq
at leading order in perturbation theory, where we recall that $2\pi N\ell^2$ is the droplet's area and
$\omega_{\text{F}}$ is the angular Fermi velocity \eqref{enek}. Thus, the rate exhibits a single absorption peak (at $\omega = \omega_{\text{F}}$) whose magnitude is independent of the direction of polarization; anisotropic droplets, to which we now turn, will change both these conclusions.

%---------------------------------------------------------------
\subsection{Microwave absorption by anisotropic droplets}
\label{semiaou}
%---------------------------------------------------------------

\begin{figure*}[t]
\centering
\includegraphics[width=0.9\textwidth]{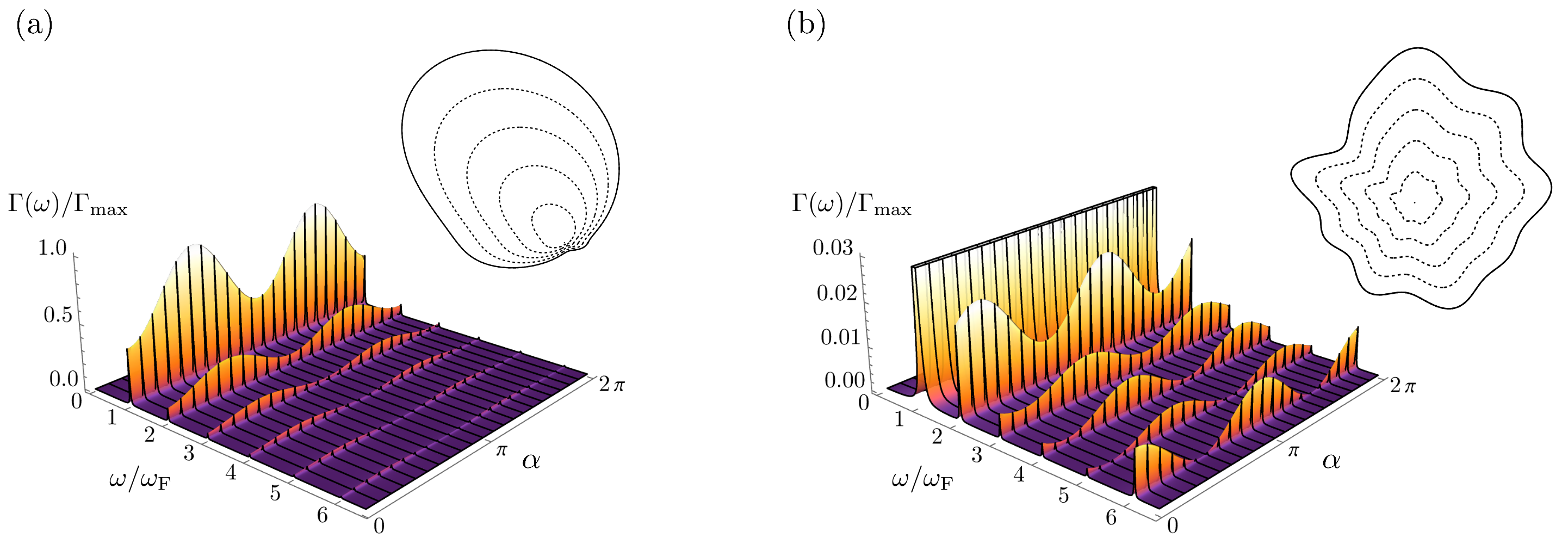}
\caption{Plots of the multiple-peaked, angle-dependent absorption spectrum \eqref{abso} for droplets (a) and (b) shown as insets, where the latter is the same as in Figs.~\ref{fiDrop} and \ref{fipott} and the dashed curves represent a few equipotentials. [As is commonplace, the delta functions in Eq.~\eqref{abso} have been replaced by Lorentzian distributions to account for the finite lifetime of quasiparticles in real systems.] Each continuous black curve displays the absorption rate $\Gamma(\omega)$ as a function of the frequency $\omega$ for a fixed value of the polarization angle $\alpha$, divided by the maximal absorption rate $\Gamma_{\text{max}}$ of the first peak. The occurrence of absorption peaks at integer multiples of the angular Fermi velocity $\omega_{\text{F}}$ is manifest. As is visible in (a), even very smooth droplets may give rise to secondary peaks $(p\geq2)$ whose magnitude is comparable to that of the first one.}
\label{fiARMABis}
\end{figure*}

Consider as before an edge-deformed potential, for which the relevant eigenstates are given by Eq.~\eqref{aspt}. Then, the one-body matrix elements needed for the transition rate \eqref{gone} read
\begin{align}
& \langle\psi_m|[x\cos(\alpha) + y\sin(\alpha)]|\psi_n\rangle \nonumber \\
& \quad \sim
\ell\sqrt{2m}\oint\frac{\dd\phii}{2\pi}\cos(\phii-\alpha)\,\ee^{\ii (n-m)f(\phii)}\,f'(\phii)^{3/2},
  \label{matell}
\end{align}
where we assumed that $m,n$ are both large with a finite difference $m-n$ and we only wrote the leading-order result in that limit. There is \i{a priori} no selection rule in Eq.~\eqref{matell}, so a state $|\psi_m\rangle$ can typically jump to any other state $|\psi_n\rangle$, with an arbitrary difference $p=n-m$. As a consequence, the many-body absorption rate $\Gamma(\omega)$ involves several distinct peaks, each labeled by $p$; at leading order in the thermodynamic limit, these peaks occur at integer multiples of $\omega_{\text{F}}$, with $p$ different one-body transitions contributing to the $p^{\text{th}}$ peak. There is thus one lowest-frequency transition at $\omega = \omega_{\text{F}}$ ($|\psi_{N-1}\rangle\to|\psi_N\rangle$), but two transitions at $\omega = 2\omega_{\text{F}}$ ($|\psi_{N-2}\rangle\to|\psi_N\rangle$ and $|\psi_{N-1}\rangle\to|\psi_{N+1}\rangle$), three transitions at $\omega = 3\omega_{\text{F}}$, and so on. All in all, the droplet's absorption rate per unit area is
\begin{align}
\frac{\Gamma(\omega)}{2\pi N\ell^2}
& \sim
\frac{q^2E^2}{2\pi\hbar^2}\,
\sum_{p=1}^{\infty}
p\,\delta\big(\omega-p\,\omega_{\text{F}}\big) \nonumber \\
&\quad\times\bigg|
\oint\frac{\dd\phii}{2\pi}\cos(\phii-\alpha)\,\ee^{\ii pf(\phii)}\,f'(\phii)^{3/2}
\bigg|^2,
  \label{abso}
\end{align}
and it manifestly depends on the polarization angle $\alpha$. As announced, this generalizes the isotropic formula \eqref{giso}, to which it reduces in the special case $f(\phii)=\phii$; it was written in Eq.~\eqref{79bis} for $\lambda(\phii)=2f'(\phii)$. Examples of absorption spectra given by Eq.~\eqref{abso} are displayed in Figs.~\ref{fiARMA} and \ref{fiARMABis}, with delta functions replaced by broader Lorentzian distributions; such a broadening typically occurs in real systems due to dissipation effects.

Eq.~\eqref{abso} predicts that secondary peaks ($p\geq2$) are typically lower than the first, although they may be of the same order of magnitude [see Fig.~\ref{fiARMABis}(a)] even for perfectly smooth potentials. In addition, the first few subleading peaks may have comparable amplitudes [see Fig.~\ref{fiARMABis}(b)], since the multiplicative factor $p$ in Eq.~\eqref{abso} increases the weight of high-frequency resonances. Their magnitude nevertheless falls off exponentially fast when $p\to\infty$, as the angular integral in Eq.~\eqref{abso} is the $p^{\text{th}}$ Fourier mode of a certain function on the circle (see Appendix~\ref{App:microwave_flower}).

In actual QH experiments, the precise shape of the trapping potential is generally unknown. Eq.~\eqref{abso} offers in this sense a promising path to reconstruct the shape of the droplet, at least partially. Indeed, the polarization of electromagnetic perturbations gives a factor $\cos(\phii-\alpha)$ in the integrand, which guarantees that the angle-dependent absorption rate at the $p^{\text{th}}$ peak is a function $A_p + B_p\cos(2\alpha+C_p)$, where the numbers $A_p$, $B_p$, and $C_p$ depend on the deformation $f(\phii)$. Each peak thus fixes (at most) three parameters entering $f(\phii)$; combining several peaks gives a fair amount of insight into the Fourier modes of the `anisotropy function' $f(\phii)-\phii$. Note in passing that just three directions of polarization are needed in order to determine the entire angular dependence of the absorption at any peak. Thus, only three waveguides with different orientations are required to measure the full angle-dependent absorption spectrum, without the need to scan every single direction individually.

\begin{figure}[t]
\centering
\includegraphics[width=0.75\columnwidth]{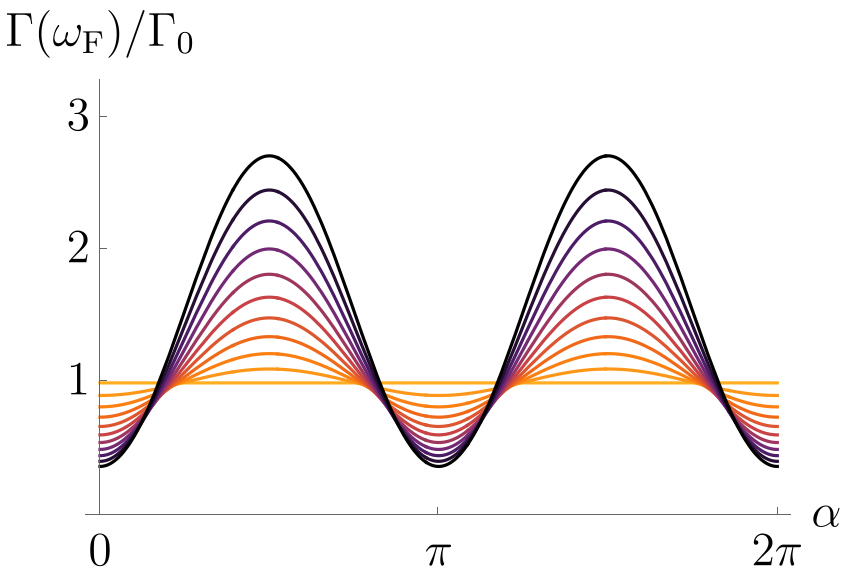}
\caption{Plot of $\Gamma(\omega_{\text{F}})/\Gamma_0=\cosh(2\lambda)-\sinh(2\lambda)\cos(2\alpha)$, describing the magnitude of the first (and only) peak in the absorption spectrum \eqref{gellip}, seen as a function of $\alpha$ for an elliptic droplet with anisotropy parameter $\lambda = 0, 0.05, \ldots, 0.5$ (orange to black). The rescaling $\Gamma_0$ contains all dimensionful quantities on the right-hand side of Eq.~\eqref{gellip} (with the delta function broadened to a Lorentzian). The case $\lambda=0$ corresponds to the circular droplet, while increasing $\lambda$ elongates the droplet along the $y$ axis and increases the absorption rate at most values of $\alpha$, except near $\alpha=0,\pi$.}
\label{fig_elliptic_abs}
\end{figure}

A key property of Eq.~\eqref{abso} is that it is consistent with the expected behavior of admittance in more familiar cases. We already mentioned this for the isotropic setup $f(\phii)=\phii$, but one can go further and consider anisotropic harmonic droplets, which were discussed in Sec.~\ref{SubSec:Elliptic_comparison}. Such droplets are obtained from isotropic harmonic traps through flower deformations \eqref{boost} of order $k=2$, \eg with $\alpha=\cosh(\lambda)$ and $\beta=\sinh(\lambda)$. Plugging such a function $f(\phii)$ into the absorption formula \eqref{abso} readily shows that a selection rule is satisfied once more: all peaks with $p\geq2$ vanish, and only the first peak ($p=1$) persists. In fact, the full angle-dependent absorption rate for an elliptic droplet is
\beq
\label{gellip}
\frac{\Gamma(\omega)}{2\pi N\ell^2}
\sim
\frac{q^2E^2}{8\pi\hbar^2}\,
\delta\big(\omega-\omega_{\text{F}}\big)
\bigl[\cosh(2\lambda) - \sinh(2\lambda)\cos(2\alpha)\bigr],
\eeq
where $\lambda$ is the deformation parameter that appears in the harmonic potential \eqref{ellipsepotential}. This can be derived either from our Eq.~\eqref{abso} or from the known exact LLL wave functions in a harmonic potential. As shown in Fig.~\ref{fig_elliptic_abs}, Eq.~\eqref{gellip} predicts that increasing the droplet's eccentricity increases its absorption rate at most values of $\alpha$, except near $\alpha = 0$ and $\pi$, where the absorption rate decreases. The intuitive explanation is that the elliptic potential \eqref{ellipsepotential} elongates the droplet along the $y$ axis, so that larger dipole moments, hence larger absorption rates, occur for $\alpha=\pm\pi/2$, while smaller rates occur near $\alpha=0,\pi$. In addition, the overall increase in the absorption rate at larger eccentricities may be viewed as an effect of the increase in the droplet's perimeter (though its area is kept constant), which agrees with the intuition that microwave absorption probes the droplet's edge dynamics.

Moving away from harmonic potentials, the next step is to consider all flower deformations \eqref{boost} with arbitrary order $k$. The corresponding potentials are both anharmonic and anisotropic (unless $k=2$), so their absorption spectra generally display infinite series of peaks. In fact, for $k=1$, all peaks are nonzero and explicitly depend on $\alpha$, while flower potentials with $k>2$ admit a (weak) selection rule of their own: the $p^{\text{th}}$ absorption peak vanishes unless $p=\pm1 \mod k$. We show in Appendix~\ref{App:microwave_flower} that this is a direct consequence of the $\mathbb{Z}_k$ symmetry of flower deformations; the same symmetry also implies (for $k>2$) that all remaining nonzero peaks are independent of the polarization angle $\alpha$, in contrast with the cases $k=1,2$ whose absorption rates are angle dependent. Actually, the precise values of absorption amplitudes \eqref{abso} can be obtained analytically for any flower deformation.

%---------------------------------------------------------------
\subsection{Nonuniform fields and subleading effects}
%---------------------------------------------------------------

It is worth noting that one is not limited to the uniform electromagnetic perturbations considered so far. Indeed, one may be interested in spatially modulated perturbations that are periodic in time, generalizing the case $W=qE[x\cos(\alpha) + y\sin(\alpha)]$ mentioned below Eq.~\eqref{gone}. One particular class of perturbations is especially well suited for the anisotropic behavior studied here, namely $W=\Lambda\,r^s \cos\bigl(n[\phii-\alpha]\bigr)$ in polar coordinates, where $s,n$ are some positive integers and $\Lambda$ is some overall scale with dimensions such that $W$ is an energy. (Incidentally, such potentials may be viewed as generators of the $W_{1+\infty}$ algebra with spin $1+s/2$ \cite{Cappelli1}.) Then, a straightforward extension of the derivation surrounding Eq.~\eqref{matell} yields the many-body absorption rate
\begin{align}
\Gamma(\omega)
&\sim
\frac{\Lambda^2\ell^{2s}(2N)^s}{2\hbar^2}
\sum_{p=1}^{\infty}
p\,\delta\big(\omega-p\,\omega_{\text{F}}\big) \nonumber \\
&\quad\times\bigg|
\oint\frac{\dd\phii}{2\pi}\cos\bigl(n[\phii-\alpha]\bigr)\,\ee^{\ii pf(\phii)}f'(\phii)^{\frac{2+s}{2}}
\bigg|^2.
  \label{gemo}
\end{align}
In contrast to Eq.~\eqref{abso}, this generally does \i{not} scale like the area of the droplet. It does, however, display a series of absorption peaks at discrete values of the frequency, though it is crucially sensitive to more Fourier modes of the deformation $f(\phii)$. For instance, one can verify that the corresponding absorption peaks generally depend on the angle $\alpha$ even for flower deformations of order $k>2$. In this sense, Eq.~\eqref{gemo} suggests that the combination of various spatially modulated rates provides an even more powerful method to reconstruct the shape of a droplet from its admittance---at the expense of being harder to realize in practice.

We end with a few words about subleading corrections that were neglected here. First, realistic absorption peaks are not delta functions as in Eqs.~\eqref{abso}--\eqref{gemo} but are broadened instead by dissipation as in Figs.~\ref{fiARMA} and \ref{fiARMABis}. Second, anharmonic terms in the energy spectrum \eqref{enek_nlin} affect both the value of the angular Fermi velocity and the dispersion relation of edge modes and edge magnetoplasmons. Both effects modify the absorption rates \eqref{abso} and \eqref{gemo}, at the very least by affecting the location and the width of absorption peaks. Other higher-order quantum corrections appear as well and can, in principle, be included similarly to those of local observables in Appendix \ref{Om-1/2corrections}. Modifications also occur due to finite-temperature effects, higher Landau levels, and interactions, none of which were taken into account in our approach. Despite these simplifications, the results presented here show how microwave absorption by anisotropic droplets can be used as a powerful probe of geometric effects in edge dynamics.

%===============================================================
\section{Conclusion and outlook}
\label{seccc}
%===============================================================

This work was devoted to a detailed study of mesoscopic droplets of noninteracting planar electrons in a strong perpendicular magnetic field, confined by any anisotropic trap with scale-invariant level curves enclosing star domains. In particular, we provided explicit formulas for the corresponding wave functions and energy spectrum, allowing us to compute the many-body density, current, and correlations of an entire droplet. The low-energy edge modes of the droplet were also shown to behave as a chiral CFT in terms of the canonical angle variable along the boundary, despite apparent inhomogeneities in terms of more na\"ive coordinates. In practice, all calculations were based on a semiclassical expansion in the magnetic length, crucially involving a WKB ansatz for holomorphic wave functions and the solution of the ensuing transport equation. The approach provides a systematic handle on perturbative quantum corrections, exemplified by our computations of subleading effects in density and current, or shape-dependent and dispersive corrections to the linear CFT spectrum.

These results pave the way for a number of applications and follow-ups. Indeed, recent advances suggest the possibility of probing local properties of QH droplets in the lab \cite{Weitz, Ahlswede, Weis, Ilani, Steele, Hashimoto2008, Hashimoto, Kamiyama:2022fqhe, Goldman1, Goldman2, Fletcher, Mukherjee:2021jjl}, both for static ground states and their dynamical edge excitations. The density \eqref{rhoo} and the current \eqref{joo} then predict observable shape-dependent effects, while the low-energy theory \eqref{lactose} predicts the ballistic propagation of local boundary disturbances with a `lab velocity' that may appear position dependent as in Eq.~\eqref{VELO}, unless one is careful to express it in canonical action-angle coordinates. Notably, we also described a realistic experiment to probe a droplet's anisotropy \i{without} local imaging, namely by measuring its microwave absorption spectrum. We showed, analytically, that the latter consists of a series of resonance peaks with a distinctive dependence on the droplet's shape and the waves' polarization; such effects are expected to be visible soon, given the high sensitivity of state-of-the-art detectors \cite{Mahoney}.

More generally, the geometry of the QH effect \cite{BradlynEtAl:2012ea, BradlynRead:2015, Klevtsov:2013iua, Ferrari:2014yba, Abanov:2014ula, Can1, Can2, Can3, Klevtsov:2015eda,Gromov:2015fda} could soon become relevant for experiments involving ultracold atoms or photonics. Our work provides a bridge between this field of mathematical physics and concrete observables in mesoscopic quantum physics. Confirming the predictions put forward here through linear response experiments, direct imaging, or measurements of edge velocity and absorption rates would be a fascinating example of many-body quantum mechanics at work.

Turning to theory, the link between our formalism and QH symmetries deserves further study: following the series of works \cite{Cappelli1, Cappelli:1992kf, Cappelli:1993ei, Cappelli:1994wb, Cappelli:1995yk, Cappelli:1996pg}, one can think of edge deformations as unitary operators acting on many-body QH states. It is then natural to wonder how these operators get composed together, since they are expected to span a Virasoro group with a nonzero central charge \cite{Cappelli:2021kxd}. More broadly, what are the operators implementing area-preserving deformations in the sense of the WKB ansatz \eqref{final}? One expects these to provide a finite (exponentiated) form of the operators studied in \cite{Cappelli1, Cappelli:1993ei, Cappelli:1994wb}, with noncommutative composition laws consistent with the geometry \eqref{pxp} of LLL physics. Similar motivations recently led to \cite{EstienneOblakSDiff} on quantum area-preserving diffeomorphisms, although the formalism developed there does not involve any LLL projection.

Most of the discussion above focused on leading-order properties, but subleading effects are sometimes crucial and deserve to be investigated in their own right. For instance, one may be interested in the irrelevant corrections of the edge field theory \eqref{lactose} mentioned at the end of Sec.~\ref{Sec:MBO:Edge_modes}, especially following the recent numerical observation \cite{Nardin_carusotto_2020} that the slow time evolution of edge density waves is governed by a nonlinear Korteweg-de Vries equation. This regime is described by \i{small} droplet deformations of the form $r^2\mapsto r^2+\alpha(\phii)$, spanning a U(1) Kac-Moody algebra whose level is sensitive to the filling fraction \cite{Cappelli1, Cappelli:1993ei, Cappelli:1994wb}. The corresponding nonlinear dynamics may then be seen as an evolution equation in an infinite-dimensional group manifold. Such a perspective is standard in geometric hydrodynamics \cite{ArnoldOrigin, ArnoldKhesin, Oblak:2020jek}, but it has only recently come to be appreciated in condensed matter physics \cite{Delacretaz_son_2022}. Our work provides a basis for considerations of this kind in the QH effect, including the possibility of inhomogeneous (position-dependent) irrelevant corrections in anisotropic traps.

Another obvious extension of this work is the fractional QH regime. In that context, no single-particle description is available, but many-body predictions such as the edge density \eqref{rhoo}, the current \eqref{joo}, or the absorption spectrum \eqref{abso} conceivably display universal geometric features that would remain true in interacting many-body ground states \cite{Geller2}. It would be thrilling to derive such predictions from the family of edge transformations studied here, either from a microscopic analysis of the Laughlin wave function or thanks to the reformulation of fractional QH states as CFT correlation functions \cite{Moore:1991ks}.

%===============================================================

\acknowledgments{We are grateful to Laurent Charles for illuminating discussions on semiclassical methods in K\"ahlerian geometric quantization. In addition, we thank Thierry Champel, Serge Florens, and Kyrylo Snizhko for discussions on semiclassical QH physics and Pascal Degiovanni and Gwendal F\`eve for pointing out possible experiments on microwave absorption. B.O.\ also thanks Mathieu Beauvillain, Nathan Goldman, and Marios Petropoulos for collaboration on related subjects. Finally, we acknowledge useful and motivating interactions with Jean Dalibard, Benoit Dou\c{c}ot, Jean-No\"el Fuchs, Marc Geiller, Gian Michele Graf, Semyon Klevtsov, Titus Neupert, Marcello Porta, Nicolas Regnault, and Paul Wiegmann.

The work of B.O.\ is supported by the European Union’s Horizon 2020 research and innovation program under the Marie Sk{\l}odowska-Curie Grant Agreement No.\ 846244. B.L.\ acknowledges funding by the European Union’s Horizon 2020 research and innovation program under Grant No.\ ERC-StG-Neupert-757867-PARATOP. P.M.\ gratefully acknowledges financial support from the Wenner-Gren Foundations under Grant No.\ WGF2019-0061. B.E.\ was supported by the ANR Grant TopO No.\ ANR-17-CE30-0013-01.}

%===============================================================
\appendix
%===============================================================

%===============================================================
\section{Isotropic droplets}
\label{appiso}
%===============================================================

Most of this work is concerned with anisotropic properties, but \i{isotropic} results provide a useful benchmark. They are simpler than their anisotropic counterparts and well known in the literature, so their properties are concisely summarized here. We begin by recalling basic aspects of the one-body energy spectrum based on the exact wave functions \eqref{phim}, then turn to many-body observables.

%---------------------------------------------------------------
\subsection{One-body spectrum}
\label{appiso:obs}
%---------------------------------------------------------------

Consider a spin-polarized 2D electron governed by the Landau Hamiltonian \eqref{h} with an isotropic confining potential $V(\bx) = V_{0}(r^2/2)$. At very strong magnetic fields, the corresponding one-body spectrum is well approximated by the solution of the LLL-projected eigenvalue equation \eqref{claim}. As the potential is isotropic, it commutes with angular momentum, so the eigenstates of $PV\!P$ are wave functions \eqref{phim} with definite angular momentum. These confirm the general near-Gaussian behavior found in Eq.~\eqref{gaussbis}: letting $|z| = \sqrt{m}+a$ with finite $a$, one finds that \eqref{phim} behaves at large $m$ as
\beq
\label{refas}
\phi_m(\bx)
=
\frac{\ee^{\ii m\phii}}{\sqrt{2\pi\ell^2}}\,
\frac{\ee^{-a^2}}{(2\pi m)^{1/4}}
\left(
1 + \frac{a^3}{3\sqrt{m}}
+ \cO(1/m)
\right),
\eeq
where we included the $m^{-1/2}$ correction for later reference. The energy $E_m$ of each state \eqref{phim} is readily found by computing the wave function $\langle z,\bar z| P V_{0}(r^2/2)P |\phi_m\rangle$, which yields the exact eigenvalue
\beq
\label{emm}
E_m
=
\langle\phi_m|V|\phi_m\rangle
=
\frac{1}{m!} \int_{0}^{\infty}\dd t\, t^m\,\ee^{-t}\, V_{0}(\ell^2t)
\eeq
in terms of the integration variable $t\equiv|z|^2$. Observe in passing that this is the value one would find from first-order perturbation theory of the full Landau Hamiltonian \eqref{h}: by construction, LLL-projected physics is only sensitive to first-order effects of the potential, while higher orders ultimately involve higher Landau levels.

Now fix an index $m\geq0$. What is the corresponding equipotential in the sense of Eq.~\eqref{ev}? To answer this in the classical limit, we let $m\gg1$ while fixing the value of $\ell^2m=\cO(1)$ and evaluate the integral \eqref{emm} by a saddle-point approximation. The outcome is
\beq
\label{emma}
E_m
=
V_{0}(\ell^2m)
+\ell^2\Omega_m
+\frac{\ell^2}{2}\Gamma_m
+\cO(\ell^4),
\eeq
where $\Omega_m$ and $\Gamma_m$ were defined in Eq.~\eqref{OmegaGamma}. This is consistent with Eqs.~\eqref{ressum_E_m} and \eqref{etop} for $\lambda(\phii)=2f'(\phii)=2$.

%---------------------------------------------------------------
\subsection{Many-body aspects}
%---------------------------------------------------------------

The sequence followed here is the same as in Sec.~\ref{semany}: we start with the density, then consider the current and the correlations close to the edge. In all cases, the edge asymptotics reproduce the formulas in Sec.~\ref{semany} for the simplest case where $f'(\phii)=1$.

\medskip
\noindent\textbf{Density.}
Let $N\gg1$ noninteracting planar electrons be subjected to the Hamiltonian \eqref{h}, with a very strong magnetic field $\bB=\dd\bA$ and a weak isotropic potential $V(\bx) = V_{0}(r^2/2)$. The ground-state wave function of this many-body system is a Slater determinant of the occupied single-particle eigenstates $\phi_0,\phi_1,\ldots,\phi_{N-1}$ given by Eq.~\eqref{phim}, each of which has a one-body density $|\phi_m(\bx)|^2$. In that specific case, the many-body density \eqref{desum} can be expressed in closed form as
\beq
\label{rhogamma}
\rho(\bx)
=
\frac{1}{2\pi\ell^2}
\frac{\Gamma(N,|z|^2)}{\Gamma(N)}
\eeq
in terms of the upper incomplete gamma function
\beq
\Gamma(N,x)
\equiv
\int_x^{\infty}\dd t\,t^{N-1}\,\ee^{-t}
=
\Gamma(N)\,\ee^{-x}\sum_{k=0}^{N-1}\frac{x^k}{k!}.
\eeq
Constancy of density in the bulk is then manifest, as is its drop to zero close to the edge $|z|=\sqrt{N}$, with an error function behavior that can be deduced from known asymptotic formulas for gamma functions \cite{Estienne:2021qqe}; this reproduces Eqs.~\eqref{ressum_rho_edge} and \eqref{rhoo} with $\lambda(\phii)=2f'(\phii)=2$.

\medskip
\noindent\textbf{Current.}
For the LLL states \eqref{phim} with definite angular momentum, each one-body current \eqref{john} is purely angular, \ie it reads $\bj_m=(\ldots)\dd\phii$. The sum \eqref{johnson} can then be evaluated in closed form owing to an exact cancellation between the contribution of the states $m$ and $m+1$, eventually leading to a current that only involves the $N^{\text{th}}$ wave function:
\begin{align}
\bJ(\bx)
&=
\sum_{m = 0}^{N - 1} \bigl|\phi_{m}(\bx)\bigr|^2 (m-|z|^2) \,\dd\phii \nonumber \\
& = \sum_{m = 0}^{N - 1}\Bigl[m\bigl| \phi_{m}(\bx) \bigr|^2 - (m+1)\bigl| \phi_{m+1}(\bx) \bigr|^2 \Bigr] \,\dd\phii \nonumber \\
\label{everybodyInDaHouse}
& = - N \bigl| \phi_{N}(\bx) \bigr|^2 \,\dd\phii.
\end{align}
It is then trivial to show that the current is localized as a Gaussian close to the edge, since this is inherited from the underlying single-particle wave function. In particular, the (leading part of the) asymptotic behavior \eqref{refas} reproduces Eqs.~\eqref{ressum_J_edge} and \eqref{joo} with $\lambda(\phii)=2f'(\phii)=2$.

\medskip
\noindent\textbf{Correlations.}
The computation of electronic correlations close to the edge is similar to that of the density. Indeed, since the many-body ground-state wave function is a Slater determinant, its two-point correlation function can be expressed as in Eqs.~\eqref{cofo} and \eqref{cofoBis}. The exact wave functions \eqref{phim} can then be used to write the correlation \eqref{cofoBis} as an incomplete gamma function (this time with a complex argument):
\beq
C(z,\bar z,w,\bar w)
=
\frac{1}{2\pi\ell^2}\,
\frac{\Gamma(N,\bar zw)}{\Gamma(N)}\,
\ee^{-(|z|^2+|w|^2)/2}\,\ee^{\bar zw}.
\eeq
It is then manifest that bulk correlations coincide with the kernel \eqref{proj} at leading order in the thermodynamic limit. As for the edge behavior, it can be extracted \eg from a steepest descent argument \cite{Estienne:2021qqe} and reproduces Eqs.~\eqref{ressum_C_edge} and \eqref{coco} with $\lambda(\phii)=2f'(\phii)=2$.

%===============================================================
\section{Semiclassical expansion of \texorpdfstring{$\boldsymbol P\boldsymbol V\!\boldsymbol P$}{PVP}}
\label{App:PVP_expansion}
%===============================================================

In this appendix, we derive Eq.~\eqref{topop} starting from Eq.~\eqref{interm}. To this end, think of $V(x,y)$ as some smooth function of $(x,y)$ whose arguments can be complexified, and change the integration variables $(x,y)$ in Eq.~\eqref{interm} to
\beq
\label{st}
s\equiv x-\tfrac{\ell}{\sqrt{2}}(z+\bar w),
\quad
t\equiv y+\tfrac{\ii\ell}{\sqrt{2}}(z-\bar w).
\eeq
In terms of $(s,t)$, the integrals in Eq.~\eqref{interm} are two line integrals in the complex plane, each along a path from $-\infty+\ii c$ to $+\infty+\ii c$, where $c$ is some irrelevant real constant (a different one for $s$ and $t$). The advantage of the change of variables \eqref{st} is to make the exponential factor in Eq.~\eqref{interm} purely Gaussian:
\begin{multline}
\label{intermm}
\!\!\!\!\!\langle z,\bar{z}|
PV\!P
|w,\bar{w}\rangle
=
\frac{1}{(2\pi\ell^2)^2}\,
\ee^{-\frac{|z-w|^2}{2}}\,
\ee^{\frac{z\bar{w}-\bar{z}w}{2}} \\
\!\!\times \int\dd s\dd t\,
V\Big(s{+}\tfrac{\ell}{\sqrt{2}}(z+\bar{w}),
t{-}\tfrac{\ii\ell}{\sqrt{2}}(z{-}\bar{w})\Big)\,
\ee^{-\frac{s^2+t^2}{2\ell^2}}.
\end{multline}
We then complexify $V$, thus replacing $V(x,y)$ by $\cV(z,\bar z)$, where $\cV(z,\bar w)$ is a function of two complex variables, holomorphic in $z$ and antiholomorphic in $w$. We can then deform \i{independently} both integration contours for $s$ and $t$ back to the real line. For small $\ell$, the Gaussian factor in Eq.~\eqref{intermm} localizes everything to $s = t = 0$. We now use our assumption of slow variation of $V(\bx)$ to Taylor expand it as
\begin{multline}
\label{expansion}
\!\! V\Big(s+\tfrac{\ell}{\sqrt{2}}(z+\bar w),t-\tfrac{\ii\ell}{\sqrt{2}}(z-\bar w)\Big) \\
\sim
\Bigl(V+\frac{s^2}{2}\der_x^2V+\frac{t^2}{2}\der_y^2V\Bigr)\Big|_{\bigl(\tfrac{\ell}{\sqrt{2}}(z+\bar w),-\tfrac{\ii\ell}{\sqrt{2}}(z-\bar w)\bigr)},
\end{multline}
where we only kept terms that give nonzero contributions to the $\cO(\ell^2)$ approximation of the integral \eqref{intermm}. Note that everything is evaluated at $(x,y)=(\tfrac{\ell}{\sqrt{2}}(z+\bar w),-\tfrac{\ii\ell}{\sqrt{2}}(z-\bar w))$; in complex coordinates, this is just the point $(z,\bar w)$, so it is simpler to write the potential as $\cV(z,\bar w)$. Plugging the expansion \eqref{expansion} into Eq.~\eqref{intermm} then yields the result \eqref{topop}.
\vspace{1em}

%===============================================================
\section{Transport equation}
\label{appexp}
%===============================================================

The goal of this appendix is to derive the real and imaginary parts of the transport equation in Eqs.~\eqref{phiprime} and \eqref{eqn}, respectively, by imposing the eigenvalue equation \eqref{claim} based on our WKB ansatz \eqref{true} in the case of edge-deformed droplets. The argument relies on expanding the energy and the potential as in Eqs.~\eqref{ev} and \eqref{topop}. It is divided in two parts. First, we use the eigenvalue equation to derive the constraint \eqref{egeqq}, and let $z$ belong to an equipotential so that the whole equation boils down to a 1D integral identity. Second, we show that the integral has a sharp saddle point in the large-$m$ limit; this allows us to rephrase the integral constraint as a first-order transport equation for the unknown function $\cn(\theta)$.

%---------------------------------------------------------------
\subsection{Evaluation along an equipotential}
%---------------------------------------------------------------

Using the wave functions \eqref{psim}--\eqref{e32} and the expansion \eqref{topop} of the potential along with the projector property $P^2=P$, the eigenvalue problem \eqref{claim} reads
\begin{multline}
\label{egeq}
0
=
\int\limits_{\RR^2}\frac{\dd^2{\bf w}}{2\pi\ell^2}\,\ee^{-\frac{|z-w|^2}{2}+\frac{z\bar{w}-\bar{z}w}{2}}\,
\Big[\Big(\cV{+}\tfrac{\ell^2}{2}\nabla^2V\Big)\Big|_{(z,\bar w)}{-}E_m\Big] \\
\times \oint\dd\theta\, \cn(\theta)\, \ee^{\ii m\theta}\,
\delta^2\Big({\bf w}-\big(F(m,\theta),G(m,\theta)\big)\Big)
\end{multline}
up to $\cO(\ell^4)$ corrections \cite{Note14}. In the case of edge-deformed traps, $\cV(z,\bar w)$ is the bicomplex potential given in Eq.~\eqref{vzw} and the delta function localizes the whole integral over ${\bf w}$ to a level curve \eqref{xypott} with $K=m$. Integrating over ${\bf w}$ and changing the integration variable from $\theta = f(\phii)$ to $\phii$ then yields Eq.~\eqref{egeqq}.

Note that the structure of Eqs.~\eqref{egeq} and \eqref{egeqq} is $0 = \ee^{-|z|^2/2}\,F(z)$ for a holomorphic function $F(z)$, so setting $F(z)=0$ on a closed curve implies $F(z)=0$ everywhere. Accordingly, we will solve Eqs.~\eqref{egeq} or \eqref{egeqq} along the equipotential \eqref{epot} by fixing $K = m$ and parametrizing
\beq
\label{zalp}
z=\sqrt{mf'(\alpha)}\,\ee^{\ii\alpha},
\quad
\alpha\in[0,2\pi).
\eeq
This ensures that all three terms in the exponential in Eq.~\eqref{egeqq} are of the same order $\cO(m)$. Then, Eq.~\eqref{egeqq} with the choice \eqref{zalp} and $\phii\equiv\alpha+\eps$ becomes
\begin{widetext}
\beq
\label{ezalp}
\begin{aligned}[b]
0
& = \int_{-\pi}^{\pi}\dd\eps\,
f'(\alpha+\eps)\,\cn(f(\alpha+\eps))
\exp\Big[\ii mf(\alpha+\eps)-\tfrac{1}{2}mf'(\alpha+\eps)+m\sqrt{f'(\alpha)f'(\alpha+\eps)}\,\ee^{-\ii\eps}\Big] \\
& \qquad\qquad \times
\Bigl[\cV\big(\sqrt{mf'(\alpha)}\,\ee^{\ii\alpha},\sqrt{mf'(\alpha+\eps)}\,\ee^{-\ii(\alpha+\eps)}\big) + \tfrac{\ell^2}{2}\nabla^2V - \Ezo_m - \ell^2 \Efo_{m} \Bigr].
\end{aligned}
\eeq
\end{widetext}
This rewriting will allow us to carry out the integral thanks to the saddle-point approximation, obtained by expanding all terms in powers of $\eps$ and leading to a differential equation for $\cn(\theta)$.

%---------------------------------------------------------------
\subsection{Saddle-point analysis and transport equation}
%---------------------------------------------------------------

The saddle-point expansion of the integral \eqref{ezalp} is cumbersome but straightforward. The strategy is to expand all factors in the integrand up to a suitable power of $\eps$, then perform the resulting integrals of the form $\int\dd\eps\,\eps^{\#} \ee^{-C\eps^2}$, where $C$ is some $f$-dependent coefficient [see Eq.~\eqref{asexp}]. The powers of $\eps$ involved are typically small, as higher powers are suppressed in the classical limit [large $m$ and $\ell^2m = \cO(1)$]. The fact that the argument of $\cn(\theta)$ also involves a factor $\eps$ eventually converts the integral into a transport equation of the form $\cn'(\theta) \propto \cn(\theta)$ [see Eq.~\eqref{transport}].

We start with Eq.~\eqref{ezalp} and first expand the exponential, then the potential with its Laplacian, and finally the simplest $f'(\phii) \cn(f(\phii))$ prefactor.
For convenience, we introduce the notation
\beq
\label{abdef}
A\equiv\frac{f''}{f'},
\qquad
B\equiv\frac{f'''}{f'}
\eeq
for combinations of derivatives of $f$ that often appear below; from now on, expressions of the form $f$ or $f'$, etc., are all implicitly evaluated at $\alpha$ unless specified otherwise [so $f\equiv f(\alpha)$, $f'\equiv f'(\alpha)$, etc.]. Note for future reference the useful relation $A'=B-A^2$.

\begin{widetext}
\medskip
\noindent\textbf{The exponential.} Using the notation \eqref{abdef}, one has
\begin{align}
\exp\Big[\ii mf(\alpha+\eps)-\tfrac{1}{2}mf'(\alpha+\eps)
+m\sqrt{f'(\alpha)f'(\alpha+\eps)}\,\ee^{-\ii\eps}\Big] \qquad\qquad\qquad\qquad\qquad\qquad\quad \nonumber \\
\sim
\ee^{\ii mf + \tfrac{1}{2}mf'}\,
\exp\left[ -\tfrac{1}{2}mf' \eps^2 \left(1+\tfrac{A^2}{4}\right) \right]
\,\left(1+mf'\eps^3\left[\tfrac{\ii}{6}-\tfrac{A}{4}-\tfrac{\ii B}{12}+\tfrac{\ii A^2}{8}-\tfrac{AB}{8}+\tfrac{A^3}{16}\right]\right),
  \label{asexp}
\end{align}
where the factor $\exp\bigl[ \ii mf + mf'/2 \bigr]$ is ultimately irrelevant for the eigenvalue equation \eqref{ezalp}, so we will not include it in what follows. The main point of Eq.~\eqref{asexp} is to exhibit the leading Gaussian behavior $\exp\bigl[ -(mf'/2)(1+A^2/4)\eps^2 \bigr]$ of the integrand, which will eventually allow us to convert Eq.~\eqref{ezalp} into a differential equation for the unknown function $\cn(\theta)$.
In fact, the same exponential factor appears in the approximately Gaussian wave function \eqref{gaussbis}.

\medskip
\noindent\textbf{The potential.}
We now turn to the expansions of the potential and of its Laplacian. As a first step, our task is to expand the potential
\begin{align}
\cV\Big(\sqrt{mf'}\,\ee^{\ii\alpha},\sqrt{mf'(\alpha+\eps)}\,\ee^{-\ii(\alpha+\eps)}\Big)
& =
V_{0}\left(\ell^2m\frac{\sqrt{f'}\sqrt{f'(\alpha+\eps)}\,\ee^{-\ii\eps}}{f'\Big(\tfrac{1}{2\ii}\log\Big[\frac{\sqrt{f'}\,\ee^{2\ii\alpha+\ii\eps}}{\sqrt{f'(\alpha+\eps)}}\Big]\Big)}\right) \nonumber \\
& \sim
V_{0}\left( \ell^2m \Big[
1 - \ii\eps \left(1+\tfrac{A^2}{4}\right)
+\eps^2 \left( -\tfrac{1}{2}+\tfrac{B}{8}-\tfrac{3A^2}{8}-\tfrac{A^3}{4\ii}-\tfrac{A^4}{16}+\tfrac{AB}{4\ii}+\tfrac{A^2B}{32} \right)
\Big] \right) \nonumber \\
& \sim
V_{0}\left(\ell^2m\right)
- \ii\ell^2m\,\eps \left( 1+\tfrac{A^2}{4} \right) V_{0}'\left(\ell^2m\right)
- \tfrac{1}{2}\ell^4m^2\eps^2 \left( 1+\tfrac{A^2}{4} \right)^2 V_{0}''\left(\ell^2m\right) \nonumber \\
& \quad\, +
\ell^2m\,\eps^2\left( -\tfrac{1}{2}+\tfrac{B}{8}-\tfrac{3A^2}{8}-\tfrac{A^3}{4\ii}-\tfrac{A^4}{16}+\tfrac{AB}{4\ii}+\tfrac{A^2B}{32} \right) V_{0}'\left(\ell^2m\right),
  \label{aspot}
\end{align}
\end{widetext}
where we used Eq.~\eqref{vzw} and the notation \eqref{abdef}. Aside from the contribution of the Laplacian, these are all the terms of the potential needed in the eigenvalue equation \eqref{ezalp} along an equipotential. As expected, they all ultimately involve the potential and its derivatives at the equipotential \eqref{epot}. For $\eps=0$, the whole expression boils down to $V_{0}(\ell^2m)$ alone.

Let us now turn to the Laplacian term. The eigenvalue equation \eqref{ezalp} requires the Laplacian evaluated at the complexified point $(z,\bar w)=\big(\sqrt{mf'(\alpha)}\,\ee^{\ii\alpha},\sqrt{mf'(\alpha+\eps)}\,\ee^{-\ii(\alpha+\eps)}\big)$. In practice, the Laplacian term is multiplied by $\ell^2$ in Eq.~\eqref{ezalp}, so we may safely set $\eps=0$ when computing it; this removes the complexification and allows us to write the Laplacian contribution in Eq.~\eqref{ezalp} as
\begin{align}
\tfrac{\ell^2}{2}\nabla^2V
& \sim
\tfrac{\ell^2}{f'}
\left( 1-\tfrac{B}{4}+\tfrac{A^2}{2} \right)
V_{0}'(\ell^2m) \nonumber \\
& \quad +
\tfrac{\ell^4m}{f'} \left(1+\tfrac{A^2}{4}\right) V_{0}''(\ell^2m),
  \label{aslap}
\end{align}
which follows from the general expression \eqref{lav} evaluated on the equipotential \eqref{epot}.

\medskip
\noindent\textbf{All together.}
Let us finally consider the very first factor on the right-hand side of Eq.~\eqref{ezalp}, namely
\beq
\label{aspref}
f'(\alpha+\eps)\,\cn(f(\alpha+\eps))
\sim
f'\cn(f)+\eps\bigl[f''\cn(f)+f'^2\cn'(f)\bigr],
\eeq
where higher powers of $\eps$ are negligible at this order. To see why they may be neglected, it is helpful to investigate the general structure of the small-$\ell$ expansion in Eq.~\eqref{ezalp}: the exponential term in Eq.~\eqref{asexp} has the form
\beq
\label{schem}
\exp[\ii mf(\ldots)]
\sim
\mathrm{const} \times \ee^{-m\Lambda\eps^2} \bigl(1+mL\eps^3\bigr)
\eeq
with $m\gg 1$ and $\Lambda,L$ some $\cO(1)$ coefficients. Similarly, the potential expansion \eqref{aspot} together with the Laplacian correction \eqref{aslap} can schematically be written as $V_{0}+\tfrac{\ell^2}{2}\nabla^2V_{0} \sim V_{0}+\ell^2W_0+G\eps+H\eps^2$, where $V_{0} \equiv V_{0}(\ell^2m)$ while $W_0$, $G$, $H$ are again some $\cO(1)$ coefficients. Finally, the expansion \eqref{aspref} of the prefactor roughly has the form
\beq
\label{schemm}
f'\cn\, \ee^{(\ldots)}
\sim
\mathrm{const} \times (f'\cn+\eps I\cn'+\eps J\cn),
\eeq
where $I,J$ are $\cO(1)$ coefficients. Putting together the schematic expressions \eqref{schem} and \eqref{schemm} and using the fact that constant (\ie $\eps$-independent) contributions are irrelevant, the eigenvalue equation \eqref{ezalp} becomes
\begin{multline}
\label{inteq}
0 = \int\dd\eps\,
\bigl(f'\cn+\eps I\cn'+\eps J\cn\bigr)
\ee^{-m\Lambda\eps^2} \bigl(1+mL\eps^3\bigr) \\
\times
\Bigl(V_{0}+\ell^2W_0+G\eps+H\eps^2 - \Ezo_m - \ell^2 \Efo_{m} \Bigr).
\end{multline}
Here, the right-hand side is a sum of integrals whose integrand has the form $\eps^n \ee^{-m\Lambda\eps^2}$.
For odd $n$, each such integral vanishes; for even $n$, it is nonzero and scales as $m^{-n/2}$. This is why only the first order in $\eps$ is needed in the expansion \eqref{aspref}: higher powers of $\eps$ would yield subleading corrections to Eq.~\eqref{inteq}, which can be consistently taken into account only by expanding the exponential, potential, and Laplacian terms up to orders in $\eps$ higher than what we did above. Here, we content ourselves with the zeroth- and first-order terms in $\ell^2$ (\ie in $1/m$). At that level of approximation, Eq.~\eqref{inteq} yields the zeroth-order statement 
\beq
\label{adgt}
V_{0} - \Ezo_m = 0
\eeq
and the first-order result
\begin{multline}
\label{scheg}
f'\cn
\Bigl[\Lambda \ell^2m (W_0 - \Efo_{m}) + \tfrac{H}{2} + \tfrac{3LG}{4\Lambda}\Bigr]
+\tfrac{G}{2} \bigl( I\cn' + J\cn \bigr)
=
0,
\end{multline}
where $\ell^2m=\cO(1)$ as before. Eq.~\eqref{adgt} confirms that the eigenvalue equation holds if $\Ezo_m = V_{0}(\ell^2m)$, \ie if the energy of the eigenstate $|\psi_m\rangle$ is that of its equipotential at leading order [recall Eq.~\eqref{ev}]. More important, Eq.~\eqref{scheg} yields a transport equation for $\cn$, whose schematic form is
\beq
\tfrac{GI}{2} \tfrac{\cn'}{\cn}
+f'
\Bigl[\Lambda \ell^2m \bigl(W_0-\Efo_m\bigr)
+\tfrac{H}{2}
+\tfrac{3LG}{4\Lambda}\Bigr]
+\tfrac{GJ}{2}
=
0. \vspace{1mm}
\eeq
We now rely on the expansions \eqref{asexp}--\eqref{aspref} to write this transport equation explicitly: using the notation \eqref{OmegaGamma} and plugging Eqs.~\eqref{asexp}--\eqref{aspref} into Eq.~\eqref{ezalp} yields the condition
\begin{widetext}
\begin{multline}
0 = \int\dd\eps\, \ee^{-\tfrac{Kf'}{2}\left(1+\tfrac{A^2}{4}\right)\eps^2}
\biggl[ 1+\eps\left(A+f'\frac{\cn'(f)}{\cn(f)}\right) \biggr]
\biggl[ 1+Kf'\eps^3\left(\tfrac{\ii}{6}-\tfrac{A}{4}-\tfrac{\ii B}{12}+\tfrac{\ii A^2}{8}-\tfrac{AB}{8}+\tfrac{A^3}{16}\right) \biggr] \\
\times
\biggl[
-\ii\ell^2K\eps\left(1+\tfrac{A^2}{4}\right)\Omega_{m}
-\tfrac{\ell^2K\eps^2}{2}\left(1+\tfrac{A^2}{4}\right)^2\Gamma_{m}
+\ell^2K\eps^2\left(-\tfrac{1}{2}+\tfrac{B}{8}-\tfrac{3A^2}{8}-\tfrac{A^3}{4\ii}-\tfrac{A^4}{16}+\tfrac{AB}{4\ii}+\tfrac{A^2B}{32}\right)\Omega_{m} \\
+\tfrac{\ell^2}{f'} \left(1-\tfrac{B}{4}+\tfrac{A^2}{2}\right)\Omega_{m}
+\tfrac{\ell^2}{f'}\left(1+\tfrac{A^2}{4}\right)\Gamma_{m}
-\ell^2\Efo_{m}\biggr],
\end{multline}
whose structure is that announced in Eq.~\eqref{inteq}, as had to be the case. What remains is to multiply all the factors in the integrand, keep track of powers of $\eps$, and integrate over $\eps$, which leads to
\beq
\label{transport}
\ii R'/R
= \left(1 + \tfrac{A^2}{4}\right)\tfrac{\Gamma_{m}}{2\Omega_{m}}
  + 1 - \tfrac{B}{4} + \tfrac{A^2}{2}
  - f'\tfrac{\Efo_{m}}{\Omega_{m}}
  - \frac{1}{1+\tfrac{A^2}{4}} \left[ \left( \tfrac{B}{8}
  + \tfrac{A^4}{16} - \tfrac{A^2B}{32} \right)
  + \ii \left( \tfrac{A}{4}+\tfrac{3A^3}{16}-\tfrac{AB}{8} \right) \right],
\eeq
where we introduced $R \equiv R(\alpha) \equiv \cn(f(\alpha))$ for simplicity. This is the transport equation for the $\cO(1)$ multiplicative factor of the WKB ansatz \eqref{true}. Its real and imaginary parts, respectively, govern the phase and norm of $\cn(f(\phii))\equiv\cN(\phii)\,\ee^{\ii\Phi(\phii)}$:
\begin{align}
-\Phi'
& =
\left(1+\tfrac{A^2}{4}\right)\tfrac{\Gamma_{m}}{2\Omega_{m}}
- f' \tfrac{\Efo_{m}}{\Omega_{m}}
+ \tfrac{1}{1+\tfrac{A^2}{4}}
  \left( 1+\tfrac{3A^2}{4}+\tfrac{A^4}{16}-\tfrac{3B}{8}-\tfrac{A^2B}{32} \right),
  \label{rtransport}\\
\cN'/\cN
& =
- \frac{1}{1+\tfrac{A^2}{4}}
  \left( \tfrac{A}{4}+\tfrac{3A^3}{16}-\tfrac{AB}{8} \right). 
  \label{itransport}
\end{align}
\end{widetext}
The identity $B=A'+A^2$ then reduces these two relations to Eqs.~\eqref{phiprime} and \eqref{eqn} in the main text.

%===============================================================
\section{Subleading contributions}
\label{Om-1/2corrections}
%===============================================================

In this appendix, we state results for the next-order correction in $m^{-1/2}$ to the leading-order wave function \eqref{aspt}, obtained by including the $\cO(1/m)$ term in the location of the saddle point in Sec.~\ref{SubSec:GWF}. As we show, this reproduces formulas that can be derived by other means in isotropic and/or harmonic traps. We also use this to deduce $\cO(1/\sqrt{N})$ corrections to the many-body density \eqref{rhoo} and current \eqref{joo}. Note that this is \i{not} a higher-order quantum correction: the latter requires higher-order terms in the transport equation in Secs.~\ref{ssegeqen} and~\ref{SubSec:GWF}, and would give rise to $\cO(1/N)$ corrections [as opposed to $\cO(1/\sqrt{N})$] for the density and current.

%---------------------------------------------------------------
\subsection{One-body wave functions}
\label{Om-1/2corrections:OB}
%---------------------------------------------------------------

Using as before the parametrization \eqref{zalpp} near the $m^{\text{th}}$ equipotential, the integral \eqref{true} has a unique saddle point at $\phii = \alpha + \delta_1/\sqrt{m} + \delta_2/m + O(m^{-3/2})$ with
\begin{subequations}
\begin{align}
\delta_1
& =
-\ii a \Bigl[ 1 - \ii \tfrac{f''}{2f'} \Bigr]^{-1}, \\
\delta_2
& =
\ii a^2 \left( \tfrac{2{-}\{f,\alpha\}}{4} {-} \ii \tfrac{f''}{2f'} \Bigl[ 1 {-} \ii \tfrac{f''}{2f'} \Bigr] \right) \Bigl[ 1 {-} \ii \tfrac{f''}{2f'} \Bigr]^{-3},
\end{align}
\end{subequations}
where $\{f,\alpha\}\equiv f'''/f'-\tfrac{3}{2}(f''/f')^2$ is the Schwarzian derivative of $f$ and all derivatives of $f$ are evaluated at $\alpha$. Repeating the saddle-point approximation of Eq.~\eqref{true}, now keeping terms of one order in $m^{-1/2}$ more than before and using that $m \gg 1$ while $a$ is finite, one finds the wave function
\begin{widetext}
\beq
\psi_m(z,\bar z)
=
\frac{1}{\sqrt{2\pi\ell^2}}\,
\frac{1}{(2\pi m)^{1/4}}\,
\ee^{\ii mf(\alpha) + \ii\Phi(\alpha)}
\frac{1}{\sqrt{\sigma(\alpha)}}
\exp \left[ -\frac{f'(\alpha) a^2}{1-\ii\frac{f''(\alpha)}{2f'(\alpha)}} \right]
\left( 1 + \frac{1}{\sqrt{m}}R_{m}(a, \alpha) + O(1/m) \right).
\label{asptNO}
\eeq
This differs from the leading-order result \eqref{aspt} by the $m^{-1/2}$ correction
\beq
\label{Rm}
R_{m}(a, \alpha)
\equiv
a R^1_{m}(\alpha) + \frac{a^3}{3!} R^3_{m}(\alpha),
\eeq
where
\begin{subequations}
\label{Rm_coeffs}
\begin{align}
& R^1_m(\alpha)
\equiv
- \frac{f'(\alpha)}{1-\ii\frac{f''(\alpha)}{2f'(\alpha)}} \Biggl[ \frac{\sigma(\alpha)^2}{2} \biggl( \frac{\Gamma_{m}}{\Omega_{m}}+\frac{1}{2}\biggr)-\frac{\Efo_{m}}{\Omega_{m}}
+ \frac{2{-}\{ f(\alpha), \alpha \}}{8 f'(\alpha)^2 \sigma(\alpha)^2} \biggl(3{-}\ii\frac{f''(\alpha)}{2f'(\alpha)}\biggr) \biggl(1{+}\ii\frac{f''(\alpha)}{2f'(\alpha)}\biggr)\Biggr], \\
& R^3_{m}(\alpha)
\equiv
f'(\alpha) \frac{2 - \{ f(\alpha), \alpha \}}{\left( 1-\ii\frac{f''(\alpha)}{2f'(\alpha)} \right)^3},
\end{align}
\end{subequations}
\end{widetext}
expressed using Eq.~\eqref{sigma} for $\sigma(\phii)$. Note that the normalization in Eq.~\eqref{asptNO} is the same as in Eq.~\eqref{aspt} since it is unaffected by the $\cO(m^{-1/2})$ correction. [This would no longer be true when including $\cO(1/m)$ corrections.]

%---------------------------------------------------------------
\subsection{Isotropic and harmonic potentials}
\label{Om-1/2corrections:Comp}
%---------------------------------------------------------------

Let us consider the one-body wave function \eqref{asptNO} for the isotropic and harmonic cases treated in Secs.~\ref{appiso:obs} and~\ref{SubSec:Elliptic_comparison}, respectively.

\medskip
\noindent\textbf{Isotropic potential.} In this case, one has $f(\alpha) = \alpha$, so $f'(\alpha) = 1 = \sigma(\alpha)$, $f''(\alpha) = 0 = \{ f(\alpha), \alpha \}$, and $\Phi(\alpha) = \text{const}$, where we used Eq.~\eqref{equan} to get $\Efo_{m} = \Omega_{m}+\Gamma_m/2$. It follows that the coefficients in Eq.~\eqref{Rm_coeffs} are $R^1_{m}(\alpha) = 0$ and $R^3_{m}(\alpha) = 2$, meaning that Eq.~\eqref{Rm} yields $R_m(a, \alpha) = a^3/3$. In conclusion,
\beq
\psi_m(z,\bar z)
=
\frac{\ee^{\ii m \alpha}}{\sqrt{2\pi\ell^2}}\,
\frac{\ee^{-a^2}}{(2\pi m)^{1/4}}
\left( 1 + \frac{a^3}{3\sqrt{m}} + O(1/m) \right)
\eeq
up to an overall constant phase. This agrees with the asymptotics \eqref{refas} of the isotropic wave function \eqref{phim}.

\medskip
\noindent\textbf{Anisotropic harmonic potential.} In this case, $f(\alpha)$ is given by Eq.~\eqref{boost} with $k=2$, $\alpha = \cosh(\lambda)$, and $\beta = \sinh(\lambda)$. One can then show, with $\theta\equiv f(\alpha)$, that
\begin{subequations}
\begin{align}
R^1_{m}(\alpha)
& = \sinh(\lambda) \frac{\cosh(\lambda)\ee^{2\ii\theta} - \sinh(\lambda)}{\bigl[\cosh(\lambda)\ee^{2\ii\theta} + \sinh(\lambda)\bigr]^2}, \\
R^3_{m}(\alpha)
& =
2 \left( \frac{\cosh(\lambda)\ee^{2\ii\theta} - \sinh(\lambda)}{\cosh(\lambda)\ee^{2\ii\theta} + \sinh(\lambda)} \right)^3.
\end{align}
\end{subequations}
Inserting these into Eq.~\eqref{Rm} yields the $m^{-1/2}$ term in Eq.~\eqref{asptNO}, which should be seen as a correction to the leading-order result \eqref{wfunc_ellipse} stated in the main text. As in the isotropic case above, the ensuing wave function agrees with the large-$m$ approximation of the \i{exact} LLL-projected eigenstates for an anisotropic harmonic potential, which can be obtained by pushing the saddle-point analysis of \cite{ForresterJancovici:1996} one order further.

%---------------------------------------------------------------
\subsection{Many-body aspects}
\label{Om-1/2corrections:MB}
%---------------------------------------------------------------

The $\cO(1/\sqrt{m})$ correction in the wave function \eqref{asptNO} induces $\cO(1/\sqrt{N})$ corrections for many-body observables. We now write these for the density and current, whose leading-order expressions were given in Eqs.~\eqref{rhoo} and \eqref{joo}, respectively.

\medskip
\noindent\textbf{Density.}
The corrected wave function \eqref{asptNO} yields a one-body probability density $|\psi_m|^2$ that differs from the earlier result \eqref{gaussbis} by terms of order $\cO(1/\sqrt{m})$. This can be plugged into the many-body density \eqref{desum}, whereupon the sum over $m$ can be converted into an integral by the Euler-Maclaurin formula. Thus,
\begin{multline}
\rho(\bx)
\sim
\frac{1}{2\pi \ell^2} \biggl[ \frac{1}{2} \erfc\biggl(\frac{\sqrt{2}}{\sigma(\phii)}a\biggr) \\
+ \frac{\ds\ee^{-{2a^2}/{\sigma(\phii)^2}}}{\sqrt{2\pi N}} \frac{\sigma(\phii)}{2} \left( A(\phii)\frac{a^2}{\sigma(\phii)^2} - B(\phii) \right) \biggr],
\end{multline}
where
\begin{subequations}
\begin{align}
\label{aadeff}
A(\phii)
& \equiv \frac{2-\{f(\phii),\phii\}}{3f'(\phii)^2\sigma(\phii)^4}
     \Bigl[ 4 - 3f'(\phii)\sigma(\phii)^2 \Bigr], \\
B(\phii)
& \equiv \frac{\Gamma_{N}}{\Omega_{N}} + \frac{1}{2}
    - \frac{2}{\sigma(\phii)^2} \left( \frac{\Efo_{N}}{\Omega_{N}} - \frac{1}{2} \right) \nonumber \\
& \quad      
    + \frac{2-\{f(\phii),\phii\}}{12f'(\phii)^2\sigma(\phii)^4}
      \Bigl[ 4 + 3f'(\phii)\sigma(\phii)^2 \Bigr].
\end{align}
\end{subequations}
Note that this is consistent with the density of \i{isotropic} droplets: when $f(\phii) = \phii$, then $\sigma(\phii) = 1$ and $\Efo_{N} = \Omega_{N} + \Gamma_{N}/2$, implying $A(\phii)=B(\phii)=2/3$, which yields
\beq
\rho
\sim
\frac{1}{2\pi\ell^2}
\biggl[
  \frac{1}{2}\text{erfc}(\sqrt{2}a)
  + \frac{\ee^{-2a^2}}{\sqrt{2\pi N}}
    \biggl(\frac{a^2}{3}-\frac{1}{3}\biggr)
\biggr].
\eeq
The latter can also be obtained from the asymptotics of the incomplete gamma function \eqref{rhogamma} near the edge.

\medskip
\noindent\textbf{Current.}
For the many-body current, the next-order correction in the wave functions \eqref{asptNO} leads to nontrivial subleading contributions similar to those of the density, but substantially more complicated. Through strenuous computations, the sum \eqref{johnson} over one-body currents yields
\beq
\label{bJ_sublead}
\bJ(r,\phii)
=
J_{\phii}(r,\phii) \,\dd \phii
+ J_{r}(r,\phii) \,\dd r
\eeq
with the angular component
\begin{multline}
\label{J_phii_sublead}
J_{\phii}(r,\phii)
\sim
- \frac{\ds\ee^{-{2a^2}/{\sigma(\phii)^2}}}{(2\pi\ell^2)^{3/2}} \\
\times
  \ell
  \left(
    \frac{\sqrt{N}}{\sigma(\phii)}
    + A(\phii) \frac{a^3}{\sigma(\phii)^3}
    - C(\phii) \frac{a}{\sigma(\phii)}
  \right)
\end{multline}
and the radial component
\begin{multline}
\label{J_r_sublead}
J_{r}(r,\phii)
\sim
- \frac{\ds\ee^{-{2a^2}/{\sigma(\phii)^2}}}{(2\pi\ell^2)^{3/2}}
  \frac{1}{\sqrt{2f'(\phii)}}
  \frac{f''(\phii)}{2f'(\phii)} \\
\times
  \left(
    \frac{1}{\sigma(\phii)}
    + \frac{A(\phii)}{\sqrt{N}} \frac{a^3}{\sigma(\phii)^3}
    - \frac{D(\phii)}{\sqrt{N}} \frac{a}{\sigma(\phii)}
  \right),
\end{multline}
where $A(\phii)$ was defined in Eq.~\eqref{aadeff} and
\begin{subequations}
\begin{align}
C(\phii)
& \equiv D(\phii) - 1 - \frac{2-\{f(\phii),\phii\}}{2f'(\phii)\sigma(\phii)^2}, \\
D(\phii)
& \equiv \frac{\Gamma_{N}}{\Omega_{N}} + \frac{1}{2}
    - \frac{2}{\sigma(\phii)^2} \left( \frac{\Efo_{N}}{\Omega_{N}} - \frac{1}{2} \right) \nonumber \\
& \quad      
    + \frac{2-\{f(\phii),\phii\}}{4f'(\phii)^2\sigma(\phii)^4}
      \left[ 4 + f'(\phii)\sigma(\phii)^2 \right].
\end{align}
\end{subequations}
Note that the dependence on $m$ in $\Omega_m$, $\Gamma_m$, and $\Efo_m$ leads to even higher-order corrections, meaning that they can safely be evaluated at $m = N$. In the special case of isotropic potentials, the corrected components \eqref{J_phii_sublead} and~\eqref{J_r_sublead} become
\beq
\label{Jphii_isotrop}
J_{\phii}
\sim
- \frac{\ds\ee^{-2a^2}}{(2\pi\ell^2)^{3/2}}
  \left(
    \ell \sqrt{N} + \frac{2\ell}{3} a^3
  \right),
\quad
J_r
=0,
\eeq
which perfectly agree with the asymptotic behavior of the many-body current \eqref{everybodyInDaHouse} upon using both the leading and subleading parts of Eq.~\eqref{refas}.

We conclude with a few remarks on the current. First, note that the aforementioned subleading differences between $J_\phii$ and $J_r$ show that the current's tangency to the droplet only holds \i{at leading order}. Second, one can verify that the one-form \eqref{bJ_sublead} satisfies
\beq
\label{bJ_div_free}
\nabla \cdot \bJ(r, \phii) = 0 + O(1/N)
\eeq
for general anisotropic traps, as should indeed be the case for the current of any energy eigenstate. In that argument, the differences in the coefficients for the $a/\sigma(\phii)$ terms in Eqs.~\eqref{J_phii_sublead} and~\eqref{J_r_sublead} conspire so that the subleading contributions to the divergence cancel. An analogous statement appears in standard WKB theory, where the transport equation implies that the probability current is divergence-free. Our result shows that this remains true here: $\cn(f(\phii))=\cN(\phii)\,\ee^{\ii\Phi(\phii)}$ satisfying Eqs.~\eqref{phiprime} and~\eqref{eqn} is consistent with the current satisfying Eq.~\eqref{bJ_div_free}.

%===============================================================
\section{Microwave absorption for flower droplets}
\label{App:microwave_flower}
%===============================================================
 
This appendix accompanies Sec.~\ref{ARMA}; it is devoted to the microwave absorption spectrum of flower-shaped droplets given by deformations \eqref{boost} of any order $k$, including the derivation of selection rules for $k>2$. Letting $\lambda$ be the anisotropy parameter, such setups are obtained by acting on an isotropic potential with a flower deformation \eqref{boost} of the form
\beq
\ee^{\ii kf(\phii)}
=
\frac{\cosh(\lambda)\ee^{\ii k\phii}+\sinh(\lambda)}{\sinh(\lambda)\ee^{\ii k\phii}+\cosh(\lambda)}.
\label{IKF}
\eeq
On top of the selection rules, we will show that the magnitude of absorption peaks can be evaluated analytically for such droplets.

Our starting point is to rewrite the angular integral in the absorption rate \eqref{abso} as a sum:
\beq
\label{aqq}
\oint\!\frac{\dd\phii}{2\pi}\!\cos(\phii{-}\alpha)\ee^{\ii pf(\phii)}f'(\phii)^{3/2}
= \frac{1}{2}(X_p \ee^{-\ii\alpha}+\overline{X_{-p}}\ee^{\ii \alpha}),
\eeq
where we defined
\begin{equation}
\label{xipp}
X_p
\equiv
\oint \frac{\dd\theta}{2\pi}\ee^{\ii p\theta}\frac{\ee^{\ii f^{-1}(\theta)}}{\sqrt{(f^{-1})'(\theta)}}
\end{equation}
in terms of the canonical angle coordinate $\theta=f(\phii)$. Thus, the coefficients $X_p$ and $\overline{X_{-p}}$ that determine the magnitude of the $p^{\text{th}}$ peak are Fourier modes of the auxiliary function
\begin{equation}
\label{auxf}
F(\theta)
\equiv
\frac{\ee^{\ii f^{-1}(\theta)}}{\sqrt{(f^{-1})'(\theta)}}
=
\sum_{p=-\infty}^{\infty}X_p \ee^{-\ii p\theta}.
\end{equation}
This readily implies the following selection rule: if $f(\phii)$ is a flower deformation of order $k$ as in Eq.~\eqref{IKF}, then
\beq
\label{wsr}
X_p = 0\quad\text{if}\;\; p\neq-1\!\!\!\mod k.
\eeq
Put differently, the $p^{\text{th}}$ absorption peak can be nonzero only if $p=\pm1\mod k$. To prove this, note first that the definition of flower deformations \eqref{IKF} is ambiguous: if $f(\phii)$ satisfies Eq.~\eqref{IKF}, then so does $f(\phii)+2\pi/k$. We fix this ambiguity by choosing $f$ smooth and such that $f(0)=0$, hence
\begin{equation}
f(\varphi) 
=
-\frac{2}{k}\arctan\left(\ee^{2\lambda}\cot(k\varphi /2)\right)+\frac{2\pi}{k} \floor*{\frac{k\varphi}{2\pi}} + \frac{\pi}{k},
\end{equation}
where $\floor*{\cdot}$ denotes the nearest lower integer part. As a consequence, the auxiliary function \eqref{auxf} satisfies $F(\theta+2\pi/k) = \ee^{2\pi\ii/k}F(\theta)$, which, in turn, implies the announced selection rule \eqref{wsr}.

A corollary of this observation is that the norm of the expression in Eq.~\eqref{aqq} is independent of $\alpha$ for any $k>2$, since at least one of $X_p = 0$ or $\overline{X_{-p}} = 0$ must be true. Thus, the uniform-field absorption rate \eqref{abso} is independent of $\alpha$ for any flower deformation beyond the elliptic case, and the peaks in the absorption rate \eqref{abso} vanish unless $p = \pm1 \mod k$. This does \i{not} mean that all peaks allowed by the selection rule \eqref{wsr} are nonzero; for instance, the case $k=2$ allows peaks for any odd $p$, but all peaks vanish in practice except $p=1$, where the absorption rate is given by Eq.~\eqref{gellip}. The other exceptional case is $k=1$, where the selection rule trivially allows all peaks to be nonzero and angle dependent---as indeed they are.

For $k>2$, the rule \eqref{wsr} turns out to give all the vanishing peaks. The intensity of the remaining nonzero peaks can be evaluated analytically. Indeed, for $p  = - 1  + nk$ with some integer $n$, one can rewrite the Fourier mode \eqref{xipp} as
\begin{align}
X_{-1+nk}
&= k \int_0^{2\pi/k} \frac{\dd\theta}{2\pi} \, \ee^{\ii n k \theta}
  \frac{\ee^{\ii [f^{-1}(\theta) - \theta]}}{\sqrt{(f^{-1})'(\theta)}} \nonumber \\
&= \int_0^{2\pi} \frac{\dd \phi }{2\pi} \, \ee^{\ii n \phi} G(\phi),
\label{xapiti}
\end{align}
where we let $\phi\equiv k\theta$ and the function
\begin{align}
G(\phi)
& \equiv \frac{\ee^{\ii [f^{-1}(\phi/k) - \phi/k]}}{\sqrt{(f^{-1})'(\phi/k)}}
\end{align}
is $2\pi$-periodic. Using the flower deformation \eqref{IKF} and the fact that its inverse takes the same form with $\lambda$ replaced by $-\lambda$, one finds
\begin{align}
G(\phi)
& = \cosh(\lambda)
      \bigl[1{-}\tanh(\lambda)\ee^{-\ii\phi}\bigr]^{\frac{k{+}2}{2k}}
      \bigl[1{-}\tanh(\lambda)\ee^{\ii\phi}\bigr]^{\frac{k{-}2}{2k}}.
\end{align}
It is then straightforward to write the Fourier modes \eqref{xapiti} in terms of hypergeometric functions: for $n \geq 0$, one has
\begin{multline}
\label{x1}
X_{-1+ nk}
= \cosh(\lambda)[-\tanh(\lambda)]^n
  \binom{\frac{1}{2} + \frac{1}{k}}{n} \\
\times\,_2F_1\left(\frac{1}{k}{-}\frac{1}{2},n{-}\frac{1}{k}{-}\frac{1}{2}; 1{+}n; \tanh^2(\lambda) \right),
\end{multline}
while, for $n < 0$,
\begin{multline}
\label{x2}
X_{-1+ nk}
= \cosh(\lambda)[-\tanh(\lambda)]^{-n}
  \binom{\frac{1}{2} - \frac{1}{k}}{-n} \\
\times\, _2F_1\left({-}\frac{1}{k}{-}\frac{1}{2},{-}n{+}\frac{1}{k}{-}\frac{1}{2};1{-}n;\tanh^2(\lambda) \right).
\end{multline}
These expressions exhibit a general pattern: the absorption rate of any nonzero peak increases when the anisotropy $\lambda$ increases. Recall from Sec.\ \ref{semiaou} that a similar behavior occurs for elliptic droplets. In fact, one can verify that Eqs.~\eqref{x1} and~\eqref{x2} reproduce the simple result \eqref{gellip} in the harmonic case $k=2$.

%===============================================================

%\addcontentsline{toc}{section}{References}
% \bibliography{aQHdroplets_references, aQHdroplets_footnotes}

\begin{thebibliography}{135}
\bibitem [{Not({\natexlab{b}})}]{Note:Reeb}%
  \BibitemOpen
  \href@noop {} {} \bibinfo {note} {{Mathematically
  speaking, our potential is assumed to be a Morse function whose Reeb graph is
  a half-line. More complicated (\eg disordered) potentials have more intricate
  Reeb graphs, but the semiclassical approach developed here is likely to apply
  to each of their graph's separate branches.}}\BibitemShut {Stop}%
\bibitem [{Not({\natexlab{h}})}]{Note:Fourier}%
  \BibitemOpen
  \href@noop {} {} \bibinfo {note} {{To be precise, the flow
  of any vector field of the form
  $[A+B\cos(k\phii)+C\sin(k\phii)]\partial_{\phii}$ consists of diffeomorphisms
  of the form \eqref{boost}. This is the sense in which `flower deformations
  are akin to Fourier modes for circle diffeomorphisms'.}}\BibitemShut {Stop}%
\bibitem [{Not({\natexlab{p}})}]{Note:Largem}%
  \BibitemOpen
  \href@noop {} {} \bibinfo {note} {{At finite $m,n$, the
  exact wave functions \eqref{phim} have the same matrix elements up to
  slightly different coefficients in front of the two Kronecker deltas, but
  this difference disappears at large $m,n$ and is therefore unimportant for
  us.}}\BibitemShut {Stop}%
\end{thebibliography}

%apsrev4-2.bst 2019-01-14 (MD) hand-edited version of apsrev4-1.bst
%Control: key (0)
%Control: author (8) initials jnrlst
%Control: editor formatted (1) identically to author
%Control: production of article title (0) allowed
%Control: page (0) single
%Control: year (1) truncated
%Control: production of eprint (0) enabled
%

%===============================================================
\end{document}